\documentclass[acmlarge]{acmart}

\usepackage{tikz}
\usepackage{amsmath}

\usepackage{filecontents}

\usepackage{booktabs} 
\usepackage{makecell}
\usepackage{multirow}
\usepackage{csquotes}
\usepackage{subcaption}
\usepackage{supertabular}

\makeatletter
\patchcmd{\csq@bquote@i}{{#6}}{{\emph{#6}}}{}{}
\makeatother

\makeatletter
\let\oldlt\longtable
\let\endoldlt\endlongtable
\def\longtable{\@ifnextchar[\longtable@i \longtable@ii}
\def\longtable@i[#1]{\begin{figure}[t]
\onecolumn
\begin{minipage}{0.5\textwidth}
\oldlt[#1]
}
\def\longtable@ii{\begin{figure}[t]
\onecolumn
\begin{minipage}{0.5\textwidth}
\oldlt
}
\def\endlongtable{\endoldlt
\end{minipage}
\twocolumn
\end{figure}}
\makeatother

\newcolumntype{L}[1]{>{\raggedright\let\newline\\\arraybackslash\hspace{0pt}}m{#1}}
\newcolumntype{R}[1]{>{\raggedleft\let\newline\\\arraybackslash\hspace{0pt}}m{#1}}
\newcolumntype{C}[1]{>{\centering\let\newline\\\arraybackslash\hspace{0pt}}m{#1}}

\newcommand{\todo}[1]{\textcolor{red}{TODO}}
\newcommand{\uten}[0]{F1}
\newcommand{\efour}[0]{F2}
\newcommand{\esix}[0]{F3}
\newcommand{\fone}[0]{F4}
\newcommand{\ftwo}[0]{F5}
\newcommand{\eeight}[0]{F6}
\newcommand{\ffive}[0]{F7}
\newcommand{\fsix}[0]{F8}
\newcommand{\fthree}[0]{F9}
\newcommand{\fseven}[0]{F10}
\newcommand{\fnine}[0]{F11}
\newcommand{\feleven}[0]{F12}

\newcommand{\pfive}[0]{P3}

\newcommand\filleddot{\,\tikz\filldraw[black] (0,0) circle (3pt);}
\newcommand\emptydot{\,\begin{tikzpicture}\draw (0,0) circle (3pt);\end{tikzpicture}}
\newcommand\halfdot{\,\begin{tikzpicture}\draw (0,0) circle (3pt);\fill (0,0) -- (3pt,0) arc (0:180:3pt) -- cycle;\end{tikzpicture}}

\setcopyright{rightsretained}
\acmJournal{IMWUT}
\acmYear{2024} \acmVolume{8} \acmNumber{1} \acmArticle{34} \acmMonth{3} \acmPrice{}\acmDOI{10.1145/3643544}

\begin{document}

\date{}

\title{Matcha: An IDE Plugin for Creating Accurate Privacy Nutrition Labels}

\author{Tianshi Li}
\affiliation{%
  \institution{Northeastern University}
  \city{Pittsburgh}
  \country{USA}}
\email{tia.li@northeastern.edu}

\author{Lorrie Faith Cranor}
\affiliation{%
  \institution{Carnegie Mellon University}
  \city{Pittsburgh}
  \country{USA}}
\email{lorrie@cmu.edu}

\author{Yuvraj Agarwal}
\affiliation{%
  \institution{Carnegie Mellon University}
  \city{Pittsburgh}
  \country{USA}}
\email{yuvraj@cs.cmu.edu}

\author{Jason I. Hong}
\affiliation{%
  \institution{Carnegie Mellon University}
  \city{Pittsburgh}
  \country{USA}}
\email{jasonh@cs.cmu.edu}
\begin{abstract}
Apple and Google introduced their versions of privacy nutrition labels to the mobile app stores to better inform users of the apps' data practices.
However, these labels are self-reported by developers and have been found to contain many inaccuracies due to misunderstandings of the label taxonomy.
In this work, we present Matcha, an IDE plugin that uses automated code analysis to help developers create accurate Google Play data safety labels.
Developers can benefit from Matcha's ability to detect user data accesses and transmissions while staying in control of the generated label by adding custom Java annotations and modifying an auto-generated XML specification.
Our evaluation with 12 developers showed that Matcha helped our participants improved the accuracy of a label they created with Google's official tool for a real-world app they developed.
We found that participants preferred Matcha for its accuracy benefits.
Drawing on Matcha, we discuss general design recommendations for developer tools used to create accurate standardized privacy notices.
\end{abstract}

\maketitle

\section{Introduction}

Privacy nutrition labels are short, uniform, machine-readable notices that allow users to learn about how their data is collected and used at a glance~\cite{kelley2009, kelley2013privacy, emami2020ask}.
Inspired by this idea, the two main mobile app stores, Apple app store and Google Play store, introduced a privacy label section in 2020 and 2022 respectively.\footnote{``App privacy details'' on the Apple app store and ``Data safety section'' on the Google Play store}
This section is displayed on the public page of each app and is self-reported by the app developers.
As of August 26, 2022, 60\% of apps on the Apple app store and 44\% of apps on the Google Play store have filled out forms to create these labels.
However, researchers~\cite{li2022understanding, li2022measurement, kollnig2022goodbye, balash2022longitudinal, zhang2022usable} and consumer advocates~\cite{iPhoneap13:online} have raised numerous concerns about these labels given their current design. One key concern revolves around the accuracy of the labels, which could fundamentally undermine the entire effort if consumers lose confidence in the stated data practices. 

Currently, developers alone are responsible for accurately 
creating the privacy label.
However, recent research found that this seemingly straightforward task was very challenging for developers~\cite{li2022understanding}.
First, developers' understanding of their app's data practices may be incorrect or incomplete due to memory errors or unexpected data collection from third-party libraries~\cite{li2018coconut, mhaidli2019we, balebako2014privacy, tahaei2022charting, balebako2014improving}.
Second, developers' misinterpretations of the label terminology could cause errors when they translate the understanding of the app to the privacy label~\cite{li2022understanding}.

Using automated program or network analysis can provide insights into an app's data usage, but it can not replace developers in this process.
The output of these analyses are not directly useful to end users and need to be converted to more high-level summaries of data practices.
However, automated techniques for detecting data flows~\cite{arzt2014flowdroid, enck2014taintdroid} and categorizing them into concepts such as data types~\cite{huang2015supor, nan2015uipicker} and purposes~\cite{wang2015using, jin2018they} inevitably suffer from imperfect accuracy. 
Furthermore, after data leaves the device, it is infeasible to determine how data is stored, shared, or repurposed without access to the software backend.
As the commercial privacy labels embrace a comprehensive design that also covers server-side data rentions and sharing, the developer's knowledge is needed to elucidate the detailed data usage and correct errors made by the automated analysis.

This paper explores a new design space for tools that can improve the accuracy of standardized privacy notices by leveraging the synergies between developers and automated analysis.
We present Matcha, a plugin for the Android Studio IDE to help developers create accurate Google Play data safety labels.
Matcha analyzes an app's codebase and provides suggestions about first-party code that accesses or transmits user data based on APIs and keywords.
Matcha also automatically detects popular third-party SDKs that collect or share user data, and helps pre-fill part of the privacy labels based on the privacy information provided by the third-party developers.
Then it asks developers to confirm or reject the suggestions by adding custom Java annotations and editing an auto-generated XML spec file for first-party and third-party data practices respectively.
Finally, Matcha uses the annotations and XML to generate a CSV file that can be imported into the developer console to create the label.

The design of Matcha is inspired by prior research on privacy-enhancing IDE plugins, in which annotations has helped increase the developer's awareness of privacy issues and reinforce best practices~\cite{li2018coconut}, facilitate the documentation of data practices~\cite{li2018coconut}, and streamline the implementation of privacy features~\cite{li2021honeysuckle}.
However, the annotation design of prior research has been relatively simple, requesting information in a more open-ended format. In contrast, the creation of privacy labels requires much more comprehensive and standardized information and can be subject to developers' misunderstandings and knowledge gaps~\cite{li2022understanding}.
With Matcha, we aim to investigate the following problem: How can we apply the annotation-based approach to help developers overcome the limitations in their capacities and create an \textit{accurate} privacy label?

The design challenge lies in how to achieve a good balance between reducing developers' burden and soliciting accurate information from them.
To achieve this goal, we first analyzed Google's data safety label to design the annotations and the XML spec that covers all the required information for generating the label.
We then conducted preliminary tests for iterative design and found that developers' overconfidence and incorrect mental model of what the plugin can or cannot do made them reject correct suggestions by the plugin.
Furthermore, we noticed that our participants generally lacked basic knowledge about Java annotations, which became another barrier to providing accurate information.
These findings informed our final design of the Matcha IDE plugin.

Creating privacy labels for an app requires significant knowledge about its implementation. 
Therefore, we evaluated Matcha with 12 developers working on their own apps.
Matcha helped 11 out of the 12 participants improve the accuracy of their data safety labels as compared to filling out forms on the Google Play developer console.
Our analysis showed that Matcha was effective in addressing errors due to misunderstanding of the task of creating data safety labels, misunderstanding of the third-party libraries' data practices, forgetfulness, and misunderstanding of code behavior. All participants favored Matcha to the baseline due to improved label accuracy, user-friendly interface, learnability, and the educational benefit of learning more about their app's data practices.
Drawing on our experiences, we discuss design recommendations for developer tools for creating accurate standardized privacy notices. 

\vspace{2mm}
We make the following contributions:
\vspace{-2mm}

\begin{itemize}
    \item 
    The design and implementation of Matcha, an IDE plugin for helping developers create accurate Google Play data safety labels.
    Our annotation-based approach to creating privacy labels the IDE plugin design can address the information overload and the misunderstandings of privacy label terms, resulting in improved accuracy of privacy labels. Our plugin and source code are available at \url{https://matcha-ide.github.io}
\item 
    Evaluation studies with Android developers working on their own real-world apps ($N=12$), demonstrating the efficacy and usability of Matcha.
    \item 
    Design recommendations for developer tools for creating accurate standardized privacy notices.
\end{itemize}

\subsection{Matcha Use Case}

The example below demonstrates the typical workflow for using Matcha to create a data safety label.
Carol needs to create a data safety label for her app.
She tries the default approach, which is to fill out forms on the Google Play developer console.
However, the task is overwhelming and she is not sure whether she has answered all the questions correctly.
Then she discovers Matcha and gives it a try.

\begin{figure*}[h]
  \centering
  \includegraphics[width=1\linewidth]{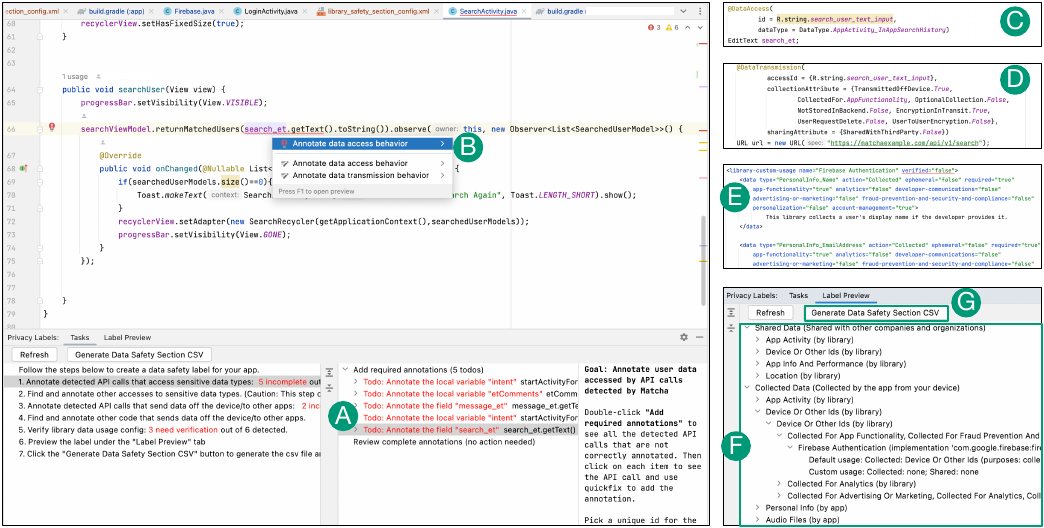}
  \caption{An overview of the main features of Matcha. Matcha detects API calls that access user data and transmit data out of the app, as well as 3rd-party SDKs that collect and share user data. Then it guides developers to confirm, refine, or reject the suggestions by adding custom Java annotations and modifying an auto-generated XML file, which account for first-party and third-party data practices respectively. Finally, Matcha generates a CSV file that can be uploaded to create the safety label.}
  \label{fig:ide-feature-overview}
\end{figure*}

Matcha analyzes the app's codebase and identifies API calls that access sensitive user data and send data off the device and asks her to add an annotation for each API call.
Carol clicks on a detected API call (\autoref{fig:ide-feature-overview}A) and is navigated to the corresponding line of code in the code editor.
She uses Matcha's quickfix (an IDE feature for repairing code issues, see \autoref{fig:ide-feature-overview}B) to add a \verb|@DataAccess| annotations (\autoref{fig:ide-feature-overview}C).
The API call accesses the search queries entered by the user, so she selects the data type ``In App Search History'' from a list of predefined options.
Similarly, she checks another detected API call that sends the search queries out of the device and also uses the quickfix feature to add a \verb|@DataTransmission| annotation that describes the ``In App Search History'' data flows from the source represented by the data access annotation to the sink represented by the transmission annotation and provides further information about why the data is being sent and how it is used after leaving the device (\autoref{fig:ide-feature-overview}D).

Finally she reviews the third-party SDKs' data practices detected by Matcha. Matcha informs her of data that is always collected and shared by the third-party SDKS integrated in her app, as well as, optional data collection and sharing which depends on her configuration of the SDKs.
Carol reviews an XML file automatically generated by Matcha which details all the potential data collection and sharing of each detected SDK and the trigger conditions (\autoref{fig:ide-feature-overview}E).
She considers the user name collection conditions of the Firebase Authentication SDK irrelevant and removes the corresponding \verb|<data>| tag.
After verifying all instances, she sets the attribute \verb|verified| to \verb|true| to indicate the completion status to Matcha.

After providing all the required information, Carol opens the ``Label Preview'' view to see the resulting label (\autoref{fig:ide-feature-overview}F).
She notices that her app both collects and shares data, while the sharing is only caused by the third-party SDKs in the app.
She also learns that data accessed and processed on device does not need to be reported as \textit{data collection} according to Google's definition of the term. 
She then clicks the ``Generate Data Safety Section CSV'' button (\autoref{fig:ide-feature-overview}G) to export the data safety label into a CSV, which she later uploads to the Google Play developer console to fulfill the requirement.

\section{Background and Related work}

\subsection{Large-scale adoption of privacy nutrition labels: Opportunities and Challenges}

Website privacy policies are notoriously long and difficult to read~\cite{mcdonald2008cost}.
To tackle the problem, researchers proposed ``privacy nutrition labels'' more than a decade ago, to offer a clear, uniform, and succinct format for disclosing data usage.
Many variants 
have been proposed for websites~\cite{kelley2009}, mobile apps~\cite{kelley2013privacy}, and IoT devices~\cite{emami2020ask}.
Prior research has shown that standardized labels can help users find information about how their data is used faster~\cite{kelley2010}, improve  comprehension of privacy practices~\cite{kelley2010}, and nudge consumers to make more privacy-conscious purchase choices~\cite{kelley2013privacy, emami2020ask}.

Apple introduced the App Privacy section to the Apple App Store in December 2020, marking the first large-scale deployment of privacy nutrition labels in real life.
Google followed with the Data Safety section, their version of privacy nutrition labels, to the Google Play Store in May 2022.
The introduction of privacy labels to the two major app stores has multiple potential benefits.
First, users can directly gain a better understanding of an app's data use~\cite{mcdonald2008cost, kelley2010}.
Second, it gives developers a systematic and structured way to disclose their data practices to end users
~\cite{li2022understanding}.
Third, the standardized and machine-readable format facilitates the research and deployment of novel formats of privacy notices to help users further synthesize, analyze, and compare app data practices~\cite{schaub2015design}.

However, researchers have identified numerous problems with these privacy labels. 
One issue is the prevalent inaccuracies in these labels.
\citet{balash2022longitudinal} found that many apps seemed likely to collect user data but 
did not declare any data collection 
in their labels.
\citet{li2022measurement} suggested that many Apple privacy labels may be outdated.
\citet{kollnig2022goodbye} found that many apps used tracking libraries and sent data to known tracking domains but reported no data collected.
\citet{xiao2022lalaine} found apps whose data flows were inconsistent with their privacy labels.
Other work focuses on usability issues.
\citet{zhang2022usable} interviewed 24 lay users about the Apple privacy labels and uncovered problems with the usability, understandability, and effectiveness of these labels.
\citet{li2022understanding} studied the usability and understandability of Apple privacy labels from the developer's perspective and identified many barriers that prevent developers from creating accurate privacy labels.
These findings suggest that the accuracy and usability problems are interdependent.

We present Matcha, an IDE plugin, to improve the accuracy of privacy labels by addressing the usability and comprehension challenges for developers~\cite{li2022understanding}.
Although we focus on Google data safety labels because they support importing labels generated by external tools, we consider the developer-in-the-loop, machine-facilitated idea generalizable to other types of privacy nutrition labels and standardized privacy notices.
We derive recommendations for designing developer tools in the same vein from in-depth studies on Matcha and discuss them at the end of the paper.

\subsection{Challenges for developers to create accurate standardized privacy notices}

Conceptually, the activity of creating a privacy nutrition label entails two steps.
The developer needs to first establish a thorough and accurate understanding of how their app handles user data~\cite{schaub2015design} and then translate it into a privacy label using the standard taxonomy.
Unfortunately, research has shown that even if developers intend to create accurate privacy labels, they often encounter significant challenges in both steps.

First, prior work has discovered reasons leading to an inaccurate understanding of the app's data practices.
\citet{li2018coconut} found that developers may lose track of changes in data practices across different versions of their app and lack knowledge about data practices in code developed by other colleagues.
Other research has revealed misunderstandings about the data practices of third-party libraries~\cite{mhaidli2019we, balebako2014privacy, tahaei2022charting, balebako2014improving}, including developers being unaware of automatic data collection by libraries they include in their apps~\cite{balebako2014privacy}. 
Third-party SDKs sometimes offer disclosures of their data practices, but developers are often unaware of these resources~\cite{mhaidli2019we, tahaei2022charting, li2022understanding}.

Second, it is difficult to synthesize data practices using standardized terms.
\citet{balebako2014your} tested both crowd workers’ and privacy experts’ ability to categorize data-sharing scenarios using a predefined taxonomy and found that participants’ understanding of the concepts in the taxonomy varied greatly, even among experts.
More specifically, \citet{li2022understanding} studied how iOS developers created privacy labels for their apps and observed that developers frequently misinterpreted terms used in the privacy label such as ``data collection''.

In this work, we introduce Matcha to address these challenges.
Matcha runs simple code analysis to identify first-party and third-party data practices, and provides scaffolding to help developers supply accurate information without spending time studying Google's definitions.
Our developer studies showed the efficacy of Matcha in enhancing the accuracy of privacy labels, and also contributed further understandings of the types of errors that Matcha helped mitigate.

\subsection{Developer tools for creating privacy notices}

\begin{table}[]
    \centering
    \begin{tabular}{p{0.22\linewidth} p{0.12\linewidth} p{0.12\linewidth} p{0.1\linewidth}  p{0.1\linewidth} p{0.1\linewidth}  p{0.1\linewidth}}
    \toprule
        Tool/Resource Name & Privacy label creation & Address \newline human Issues & Automated analysis & 3rd-party info & IDE \newline integration & Incremental update \\
    \midrule
        Honeysuckle~\cite{li2021honeysuckle} & \emptydot & \emptydot & \halfdot & \emptydot & \filleddot & \filleddot \\ 
        Official Web Forms & \filleddot & \emptydot & \emptydot & \emptydot & \emptydot & \emptydot \\
        Apple Privacy Manifest & \halfdot & \halfdot & \emptydot & \filleddot & \filleddot & \emptydot \\
        Privacy Label Wiz~\cite{gardner2022helping} & \filleddot & \halfdot & \halfdot & \halfdot & \emptydot & \emptydot \\
        Privado.ai & \filleddot & \halfdot & \halfdot & \filleddot & \emptydot & \emptydot \\
    \midrule
        Matcha & \filleddot & \filleddot & \halfdot & \filleddot & \filleddot & \filleddot \\
    \bottomrule
    \end{tabular}
    \caption{Summary of Matcha and other tools for creating privacy notices. We compare these tools along the following dimensions: Whether the tool helps with privacy label creation (\textit{Privacy label creation}); Whether the tool helps address human-related issues (e.g., infomation overload, misunderstanding of privacy label terms) that can cause inaccurate privacy labels? (\textit{Address human issues}); Whether the tool automatically detects data practices (\textit{Automated analysis}); Whether the tool provides information for filling out privacy labels for third-party SDKs used in the app (\textit{3rd-party info}); Whether the tool is integrated within the development environment so developers can have more context information and potentially work on the tasks during the development process (\textit{IDE integration}); Whether the tool supports incremental update of the privacy notice/label rather than requiring developers to scan the entire codebase from scratch every time (\textit{Incremental update}). \protect\filleddot: fully supported; \protect\halfdot: partially supported; \protect\emptydot: not supported}
    \label{tab:summary_of_tools_for_creating_notices}
\end{table}

Some existing tools can help developers audit their data practices and indirectly help developers create privacy notices.
Prior research has built information flow analyzers designed to detect malicious or unwanted information leaks from an app \cite{gordon2015information,arzt2014flowdroid,octeau2013effective,li2014know, enck2014taintdroid}.
Google and Apple have introduced similar support in recent releases of their systems, such as the data access auditing APIs introduced in Android 11 for helping developers identify unexpected data accesses.
However, most developers are unaware of and rarely use these expert features.
Furthermore, these tools are not immediately useful for privacy label creation because it is difficult to align the detection results with the types of information that a privacy label needs.

Some tools directly help developers create privacy notices and privacy labels.
Key examples are summarized in \autoref{tab:summary_of_tools_for_creating_notices} and compared with Matcha among several dimensions.
Notably, while some tools use program analysis to help identify data practices, none can fully automate this process, and all require developer input.
The privacy labels of both iOS and Android include server-side data retention practices, data usage purposes, and complex exemption rules.
Existing program analysis techniques, which focus on client-side data practices, fall short of detecting these types of information.
Below we further introduce these tools and compare them with Matcha.

Before privacy nutrition labels are widely adopted, researchers have designed tools that leverage program analysis techniques to improve the creation of various types of privacy notices, such as privacy policies~\cite{zimmeck2021privacyflash, li2018coconut} and in-app privacy notices~\cite{li2021honeysuckle}.
These tools can not generate a privacy nutrition label, which requires a more comprehensive summary of the app's data practices in a predefined taxonomy that often does not match developers' intuitive understanding.

Our work is directly inspired by and builds upon Coconut~\cite{li2018coconut} and Honeysuckle~\cite{li2021honeysuckle}, which are IDE plugins that detect sensitive data use and prompt developers to add annotations. 
Prior research has shown that adding annotations helped developers disclose more data practices when the privacy notice is written in a free form~\cite{li2018coconut} and implement contextualized privacy UIs more efficiently~\cite{li2021honeysuckle}.
However, the complex and standardized design poses more challenges to creating accurate privacy disclosures due to developers' misunderstanding and knowledge gaps~\cite{li2022understanding}.
Towards this end, we made substantial changes to the annotation design than prior work.
The Matcha annotation design breaks down the disclosure required by privacy labels into fine-grained, easy-to-understand data practices, as embodied by the annotation fields.
Our work for the first time shows that developers are able to add annotations based on code analysis results to provide precise information for creating the privacy labels. 

With the introduction of the Apple and Google privacy labels, official tools are provided for creating the privacy labels in the format of web forms.
These tools generate privacy labels based solely on developers' input and are subject to errors due to developers' misunderstanding of privacy label concepts and lack of knowledge about third-party SDKs' data practices.
Matcha significantly address these issues by leveraging the synergies between annotation-based developer input and automated code analysis, as well as informing developers of the data practices of the detected third-party SDKs.
In April 2023, Apple announced ``Privacy Manifest'', which is a property list that developers need to fill out in Xcode to describe the types of data collected by their app using the same taxonomy as privacy labels.
Third-party SDKs need to provide their own privacy manifest files.
Privacy manifests can be used as a reference to create privacy labels, potentially increasing developers' awareness of third-party SDKs. However, the fact that it is a separate requirement creates a barrier to adoption for privacy label creation.

There are other third-party tools for creating privacy nutrition labels for iOS or Android apps that use automated program analysis.
For example, Privado.ai\footnote{\url{https://www.privado.ai/data-safety-report}} is a commercial tool for creating a Google Play data safety label.
\citet{gardner2022helping} presents Privacy Label Wiz, which is a web-based tool for creating an Apple privacy label and reports on the preliminary feedback from developers.
Both Privado.ai and Privacy Label Wiz can detect potential data types and use a wizard-like interface to guide developers in providing additional information for generating privacy labels.
Both tools can detect third-party SDKs, and only Privado.ai can automatically fill out data practices of third-party SDKs.
However, these tools were not designed to address the information overload or the privacy label term misunderstanding issues, as developers are still expected to read all the text-based guidelines and term definitions on their own.
Conversely, Matcha has more design considerations like the annotation design, step-by-step task guidance, and quick-fixes to walk developers through this process and overcome these challenges.
Another difference is that these tools show the code analysis results with limited contexts (i.e., no code snippets or only short code snippets), while the IDE integration of Matcha allows for ease and flexibility of reviewing the code context, which helps developers recall the context and provide accurate privacy label information.
Furthermore, the design of Matcha has also taken the maintenance needs into account.
The use of annotations allow Matcha to efficiently solicit more fine-grained information in context.
This enables the incremental update of privacy labels.
Specifically, the developer only needs to modify the annotations around the code that has changed in the new version, and then regenerate the label.
This design eliminates the need to run the tools to scan the codebase and answer all the questions again from scratch, potentially addressing the issues of missing updates to the privacy labels~\cite{li2022measurement}.

In addition to the system design difference, our work also make research contributions by thoroughly evaluating our system and synthesizing design knowledge.
We are the first to conduct in-depth studies to show that our tool (Matcha) can improve the accuracy of the privacy label for real-world apps created by the app developers.
Our findings provide insights that can inform the design of future developer support for creating standardized privacy notices.

\section{Matcha Design and Implementation}

In this section, we present our design goals, how our design fulfills these goals, improvements to the tool design based on preliminary studies, and our final design and implementation.

\subsection{Design goals}

We drew upon prior literature to inform our design goals.
\citet{li2022understanding} discovered that one obstacle to creating accurate Apple privacy labels was that developers needed to process a large amount of new information, including lengthy and complex definitions of terms like data collection. 
The similarities between the Apple and Google label filling process suggest that Android developers may also suffer from information overload issues.
We argue that developers will benefit from more scaffolding, which leads to our first design goal:

\begin{description}
\item[D1] The privacy label questions should be deconstructed and situated within the context in which developers handle the specific code that deals with user data.
\end{description}

Developers know how their apps work, but prior research suggests they suffer from forgetfulness~\cite{li2018coconut}, lack of knowledge about other team members' code~\cite{li2018coconut} and third-party SDK's data practices~\cite{balebako2014privacy}, and misinterpretations of terms in privacy labels~\cite{balebako2014your, li2022understanding}.
Hence, we set the second design goal:

\begin{description}
\item[D2] The tool should help developers overcome limitations in their ability to create accurate privacy labels.
\end{description}

Code analysis can be used to identify some data practices that need to be reported in the data safety label. 
However, it still has limitations. 
First, it mainly analyzes data practices within the client app and cannot answer how data is used after it leaves the device.
Second, although algorithms exist to infer purposes of data use~\cite{wang2015using, jin2018they} and data types that are not tied to a specific API~\cite{huang2015supor, nan2015uipicker}, they are not always accurate.
This suggests that automated code analysis cannot be relied on completely, which leads to our last design goal.

\begin{description}
\item[D3] The tool should give developers control to refine or reject the automated analyses when they are inaccurate.
\end{description}

To satisfy \textbf{D1}, we designed Matcha as an IDE plugin.
We adopted the idea of using annotations to document data practices in code from prior work~\cite{li2018coconut, li2021honeysuckle}, allowing developers to contribute their knowledge of the app's data practices by adding annotations.
We divide the design of Matcha into two parts.
In the first part, we focus on determining what types of information to solicit from developers and in what format.
In the second part, we focus on the interaction design to help developers provide accurate information about their data use.

\subsection{Developer Input Design}

Prior work suggested that developers lack the time and ability to comprehend the label terms~\cite{li2022understanding}.
To address this issue, we designed the specific code format of developer input as a scaffolding for providing all the required information accurately.

\subsubsection{Annotations for explicit data flows within apps}

\begin{figure}
    \centering
    \includegraphics[scale=0.8]{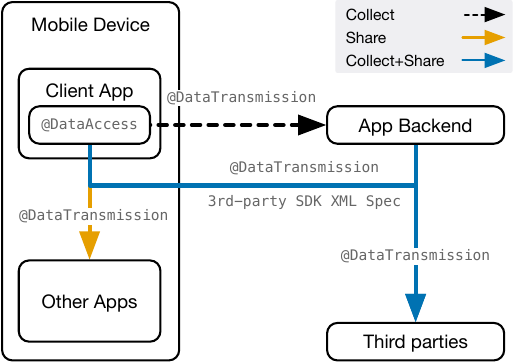}
    \caption{An illustration of the mapping between the developer input (i.e., annotations and the SDK XML spec) and Google's definitions of data collection and sharing.
    By asking developers to add data access and transmission annotations rather than directly deal with the label-specific concepts, Matcha reduces errors due to misunderstanding of the label terms while keeping developers in control of the label.}
    \label{fig:annotation-dsl-mapping}
\end{figure}

We design the \verb|@DataAccess| and \verb|@DataTransmission| annotations to indicate where data is accessed on device and transmitted to other apps or off the device, namely the sources and sinks of data flows.
\autoref{fig:annotation-dsl-mapping} shows how Matcha translates the \textit{data access} and \textit{transmission} behaviors to the label terms \textit{data collection} and \textit{sharing}, which helps address the misinterpretations of the terms.
Each \verb|@DataAccess| describes the data types accessed by the app, and each \verb|@DataTransmission| contains two sets of attributes covering the data use purposes and the special cases and exemptions of collection and sharing defined by Google. 
Developers can use \verb|@NotPersonalDataAccess| and \verb|@NotPersonalDataTransmission| to indicate no data is accessed or transmitted. 
\autoref{tab:annotation-overview} presents the annotation design.

\begin{table*}[]
    \centering
    \caption{The table shows the four annotations we designed for the task and their field members that hold different types of information needed for the label. More details about how \texttt{@DataTransmission} handles collection and sharing are in \autoref{sec:annotation-design-details}.}
    \begin{tabular}{p{0.3\linewidth} p{0.15\linewidth} p{0.5\linewidth}}
    \toprule
    Annotation & Fields & Note \\
    \midrule
        \multirow{2}{2.3cm}{\texttt{@DataAccess}} & id & A unique ID defined by the developer to refer to this access when later annotating data transmissions.\\
        & dataType & A list of Enum values of predefined data types accessed by the app and held in the annotated variable\\
        \midrule
        \texttt{@NotPersonalDataAccess} & -- & For explicitly indicating no personal data is accessed here. It is useful for dismissing an irrelevant data access suggestion. \\
        \midrule
        \multirow{3}{2.3cm}{\texttt{@DataTransmission}} & accessId &  A list of IDs indicating where the transmitted data is originally accessed. The IDs are previously defined in \texttt{@DataAccess}.\\
        & collectionAttribute & A list of Enum values of collection-related information. \\ 
        & sharingAttribute & A list of Enum values of sharing-related information. \\
        \midrule
        \texttt{@NotPersonalDataTransmission} & -- & For indicating no personal data is transmitted out here. It is useful for dismissing an irrelevant data transmission suggestion. \\
    \bottomrule
    \end{tabular}
    \label{tab:annotation-overview}
\end{table*}

\subsubsection{XML spec for implicit data flows caused by SDKs}
A library's data practices may depend on how the app uses it.  For example, in the motivating example the Firebase Authentication library only collects the user's display name if the developer provides it. 
Since much of the configuration of library data use happens outside of the app, 
it is hard to determine a proper location for the annotation and check if the required annotation is added.
Therefore, we design an XML file to allow developers to adjust the label based on their use of the library.
In the XML, Matcha generates a \verb|<library-custom-usage>| tag for each library and populates it with \verb|<data>| tags that contain the collection and sharing conditions.
Developers can either keep or remove each data tag based on the condition, and set the \verb|verified| attribute to \verb|true| to mark the configuration of a certain library as complete.

\subsection{Preliminary Tests for Iterative Design}
\label{sec:pilot-tests}

To achieve \textbf{D2}, we offer suggestions for data access and transmission and quickfixes for adding annotations based on code analysis.
To achieve \textbf{D3}, we let developers have the final say, namely, they can ignore any suggestions, and the label creation only relies on the annotations and the XML spec that they can modify. 
This raises a question: \textit{Are developers capable of correctly comprehending the suggestions and building on them to create an accurate data safety label?}
We iteratively improved the design to achieve this goal by conducting IRB-approved preliminary tests with five developers on an initial prototype with basic support for adding annotations and editing the XML spec.
The participants first created the label for their apps by Google's tool and then using Matcha.
The interview script can be found in \autoref{sec:interview-script}.
One researcher conducted a thematic analysis of the interview transcripts.
Below, we summarize issues that emerged from the studies and how they informed the improvement of our system design.

\subsubsection{Ignoring unexpected suggestions}
\label{sec:pilot-tests-overconfidence}
Some developers appeared to be affected by confirmation bias~\cite{chattopadhyay2020tale}, ignoring suggestions that did not match their expectations.
For example, \pfive{}'s app had a feature for sharing the user's high score, the game screenshot, and a short message provided by the user.
This data falls under the ``App activity'' data category per Google's definition.
However, \pfive{} quickly added a \verb|@NotPersonalDataAccess| annotation to dismiss Matcha's suggestion.
He explained that \blockquote{I don't think high score is personal info}.
This example shows that developers tended to place more trust in their understandings than the system's suggestion.
This created obstacles to correcting the developer's misunderstandings.
To address this problem, we consulted the guidelines for human-AI interaction~\cite{amershi2019guidelines} as building trust is the main goal of constructing efficient AI-infused systems.
Specifically, we added proactive and contextualized guidance to help developers establish a clearer mental model of what Matcha can do and how Matcha's suggestions work.

\subsubsection{Difficulty of handling Java annotations}
Our participants faced  challenges in adding the annotations due to unfamiliarity with the syntax.
For example, Java annotations can only be added to specific code elements, such as a variable declaration.
This troubled developers when they needed to manually add the annotation.
Another example is when developers declare multiple variables altogether (e.g., \verb|EditText nameText, ageText;|) they can not add different annotations to each variable.
During our preliminary tests, we found that these seemingly trivial issues with annotations greatly hindered our participants' abilities to use our tool. Therefore, we optimized the support for adding annotations in the final version of Matcha to help developers automatically trace where to add the annotations and reformat their code.

\subsubsection{Error-prone direct editing of annotations}
Some errors were introduced when developers edited annotations.
Since our participants were first-time users of Matcha and unfamiliar with the Google data safety label design, they felt overwhelmed by the number of fields they needed to manually complete and the number of predefined values they can select from.
Although direct editing may be more efficient for expert users, it was too error-prone and confusing for novices.
Therefore, we provided a dialog for guiding the developer to fill out all the required information to generate the annotation.

\subsection{Final Design of the IDE Plugin}
\label{sec:final-interaction-design}
We present the final design of Matcha, including the five tasks for creating the label and the main supporting features.

\subsubsection{A five-step process}

The Matcha label creation process consists of five steps. 
The first two are for adding data access annotations and the next two are for transmission annotations.
In the first and the third steps, Matcha guides developers to do a precise API-based search, in which they annotate all the detected API calls that potentially access user data and send the data out of the app.
In the second and the fourth steps, Matcha uses fuzzy keyword-based search to help developers detect more data types that are collected by the app but have not been annotated in the previous step.
Adding annotations is voluntary for the second and the fourth steps, which means the developer only needs to add an annotation when they find the detected keywords relevant to the access to, and transmission of, user data.
In the last step, the developer modifies the auto-generated XML spec file to adapt it based on their usage of the library.

\begin{figure*}
\centering
\begin{subfigure}[b]{0.42\linewidth}
\includegraphics{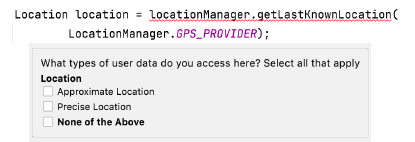}
\caption{Data type options customized based on the API call.}
\label{fig:access-quickfix-customization}
\end{subfigure}
\hspace{0.05\linewidth}
\begin{subfigure}[b]{0.5\linewidth}
\includegraphics{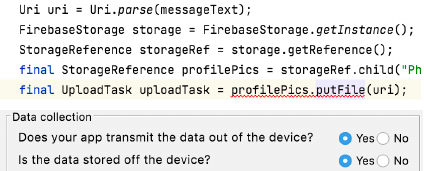}
\caption{Transmission and storage pre-checked based on the API call.}
\label{fig:transmission-quickfix-customization}
\end{subfigure}
\caption{Examples of quickfix dialogs customized based on the API calls to avoid errors and improve learnability and usability.}
\end{figure*}

\subsubsection{Quickfixes for adding annotations}
Matcha offers quickfixes to aid in the annotation creation (see \autoref{fig:ide-feature-overview}).
The quickfix locates which variable to annotate for detected API calls.
The developers can also use the quickfix to add additional annotations for any variables.
The quickfix dialog can narrow down or pre-select options based on the detected API call.
For example, for \verb|LocationManager.getLastKnownLocation|, the available choices are only approximate location, precise location, and none of the above (\autoref{fig:access-quickfix-customization}).
For a Firebase Cloud Storage API, the ``data being transmitted off the device'' and ``data being stored'' options are automatically checked (\autoref{fig:transmission-quickfix-customization}).
We note that users can still modify the annotations in any way they want, while using the dialog to constrain and validate the developer's input could both reduce the chance of accidentally making errors and make the tool easier to learn and use~\cite{norman2013design}.

\begin{figure*}
\centering
\begin{subfigure}[b]{0.45\linewidth}
\includegraphics{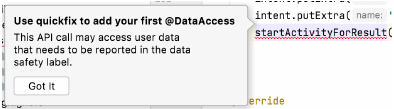}
\caption{A tooltip suggesting next action.}
\label{fig:got-it-tooltip-instruction}
\end{subfigure}
\hspace{0.05\linewidth}
\begin{subfigure}[b]{0.45\linewidth}
\includegraphics{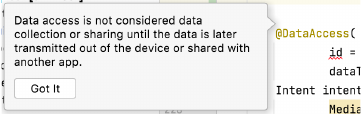}
\caption{A tooltip explaining the generated annotation.}
\label{fig:got-it-tooltip-education}
\end{subfigure}
\caption{Examples of the contextualized, proactive tooltips to offer just-in-time instructions and education about the annotations.}
\label{fig:got-it-tooltip-examples}
\end{figure*}

\subsubsection{Contextualized and proactive guidance}
To help developers understand what Matcha can do and how Matcha works, we designed just-in-time tooltips that pop up next to the related code when a certain type of Matcha suggestions is shown for the first time to give instructions on the expected actions (\autoref{fig:got-it-tooltip-instruction}).
Matcha also provides tooltips that are only informational, such as explaining what the annotations have to do with the creation of the safety label (\autoref{fig:got-it-tooltip-education}).
In addition, Matcha offers a systematic introduction of each step in a help panel (see \autoref{fig:ide-feature-overview}).

\subsubsection{Label preview}
To help developers better understand how the data safety label is generated, we design a label preview panel (see \autoref{fig:ide-feature-overview}).
For each data type that is collected or shared, a note of ``by library,'' ``by app,'' or ``by app and library'' is provided, indicating the source of this data collection or sharing.
After expanding each data type, it will show further information like the data collection or sharing purposes, as well as the related code links.
These links allow the developer to check which annotations, custom library usage records in the XML file, or third-party libraries that always collect user data led to the generation of this particular data safety label entry.

\subsection{Matcha System Implementation}
The Matcha IDE plugin was developed using the IntelliJ Platform SDK\footnote{\url{https://plugins.jetbrains.com/docs/intellij/welcome.html}}.
We have released the plugin on JetBrain's official plugin store and open sourced it on GitHub\footnote{\url{https://matcha-ide.github.io}}. 

Our API-based detection was built upon the code analysis subsystem of Coconut~\cite{li2018coconut}.
For data access, we augmented the Coconut API list with APIs from the official guidelines for this task\footnote{\url{https://developer.android.com/guide/topics/data/collect-share}}.
For transmission, we kept the APIs about network requests from Coconut and added on-device sharing API calls for the sharing-only condition.
Our final API list contains 91 data access APIs and 45 data transmission APIs.

We implemented the keyword search using the IntelliJ SDK's \verb|findManager.findInProject| API.
Our keyword list comes from three sources: (1) permissions mapped with specific data types in Google's official guideline\footnote{\url{https://developer.android.com/guide/topics/data/collect-share}}, (2) keywords extracted from Google's data type definitions\footnote{\url{https://support.google.com/googleplay/android-developer/answer/10787469?hl=en\#data_types}}, and (3) keywords extracted from open-sourced Android apps using a TF-IDF algorithm.
For the third approach, we selected 75 apps by searching for recently updated GitHub repos (in July 2020) that contain a Google Play link and declare dangerous permissions in the manifest.
To identify the keyword candidates for the data types to be reported in the safety label, we tokenized the Java files from these projects and split the variables, treated each file as a document, and calculated the TF-IDF of each word per document.
We then selected files containing the data access API calls and ranked words appearing in them based on the average TF-IDF values.
Finally, a researcher reviewed the top results and incorporated words related to the data type into the final list.
Matcha's keyword search feature is case-insensitive. Our final version covers 180 unique keywords (see \autoref{sec:implementation-details} for details).

Matcha scans the app's build.gradle file to detect third-party SDKs and fills out the required data collection and sharing practices in the generated label.
Because privacy labels (for both iOS and Android) heavily involve server-side data practices (such as retention and sharing), it is infeasible to automate the analysis of third-party SDKs by running static/dynamic program analysis.
Therefore, the auto-generated privacy label information for third-party SDKs was obtained by one researcher manually coding the privacy label filling guidelines provided by third-party SDK developers (e.g., the documentation of the Firebase SDKs.\footnote{\url{https://firebase.google.com/docs/android/play-data-disclosure}}).
We covered popular commercial SDKs included in the Google Play SDK Index\footnote{\url{https://developer.android.com/distribute/sdk-index}}.
We also added SDKs developed by Google that provided privacy label guidelines.
The SDK list is dynamically loaded every time our plugin loads by requesting a JSON file hosted remotely, allowing the list to be updated without updating the entire plugin.
Our final version for the study covers 58 unique third-party SDKs (see \autoref{sec:implementation-details} for details).
Prior research~\cite{chitkara2017does} has shown that over half of the sensitive data access by third-party libraries are from the most popular 30 libraries, which suggests that our current implementation could help developers cover a large portion of the data practices due to third-party SDK in their privacy labels.

For the SDKs that involve optional data collection and sharing practices, Matcha generates XML code to allow for further customization by the developer.
For each detected SDK, Matcha generates a \texttt{<library-custom-usage>} tag with the initial value of the \texttt{verified} attribute set to false.
Then it inserts \texttt{<data>} tags under the \texttt{<library-custom-usage>} to represent the data collection and sharing instances derived from the guidelines of the SDK.

\section{Matcha Evaluation}
\label{sec:formal-eval-studies}

\subsection{Study Design Considerations}

Evaluating interventions for improving the accuracy of a privacy label is difficult.
Unlike many developer tools that can be evaluated with uniform tasks in well-controlled settings~\cite{li2018coconut, li2021honeysuckle, drosos2020wrex, wang2022diff, wang2022documentation}, a tool for creating the privacy label must be evaluated by developers who have adequate knowledge about the app.
Otherwise, it becomes hard to eliminate the impact of lacking familiarity of the app. 
One potential method is to ask participants to develop an app with specific data practices and then create the label.
However, even developing a small app can cost thousands to tens of thousands of dollars, making it too costly for research. Hence, we chose to ask participants to work on a real-world app they developed.

However, asking participants to work on their own apps is also challenging. 
First, we cannot obtain the ground truth of the data practices of these apps to verify the developers' answers.
Second, since participants works on different apps, we cannot directly compare their performance, making a between-subjects study design unsuitable.
\label{sec:different-project-limitations}
Third, it is hard to recruit a large sample of participants who not only have developed an Android app but are also willing to install a plugin to analyze their code and let researchers look at their code.

Given these challenges, we conducted within-subjects, mixed-methods studies to gain quantitative and qualitative insights for this question: \textit{How effective is Matcha in correcting the errors in a privacy label created by Google's official tool?}
We observed how developers created the label using Google's tool and Matcha.
We asked them to compare the two labels to measure the change in accuracy in the absence of absolute ground truth.
The realistic setting placed higher requirement on our tool to work with arbitrary apps~\cite{guo2021ten}.

\subsection{Participants}

12 developers participated in our study. We coded the data iteratively and stopped recruiting after reaching saturation in our qualitative analysis~\cite{hennink2017code}.
This sample size is consistent with evaluation studies of novel programming tools in prior work ~\cite{liu2022wigglite, li2021honeysuckle, drosos2020wrex, zheng2022telling, zhang2020interactive}.
Most participants were recruited from freelancer websites, including eight from Freelancer\footnote{\url{https://www.freelancer.com}} and one from Upwork.\footnote{\url{https://www.upwork.com}}
The other three signed up after seeing advertisements on Twitter, Slack groups, or from personal connections.
Our pre-screening survey asked for an Android app they developed and their role(s) in the development process, and had quiz questions about Android development.

Our participants came from eight countries and developed the apps either as part of their job, a hobby, or for a course project.
Nine out of the 12 participants had experience in publishing apps on the Google Play store, though some chose not to use Google Play apps for our study due to NDA restrictions.
Most of our participants have low familiarity with the task of creating privacy labels.
Nine out of the 12 participants had heard about the Google privacy label as a developer or a user, while only three have created one.
We also asked about their familiarity of iOS privacy labels, and only four have heard about it and two have created one.
Our sample included six Google Play apps, with the most popular one having over one million downloads.
We append details about each participant in \autoref{sec:participant-overview}.

\subsection{Study Procedure}

We started the study by briefing the participant on the study goals and obtained their consent for audio and screen recording.
Before the main tasks, we first gave a brief introduction to the Google Play data safety label.
Then we asked the participants to introduce the app they had selected.

In the first task, the participant created the label for the selected app by filling out forms on the developer console.
Participants logged into the console using an account we created for the study.
We asked them to handle this task as they normally would and encouraged them to use any resources they would normally consult, except for the app's current label if available.
In the second task, the participant created the label using Matcha.
We first helped them download and install the plugin and then asked them to watch a short tutorial video (2 minutes 42 seconds) before working on the task.
If they were not sure about how to fill in certain information, we asked them to answer based on their best understanding and explain their rationale.
After creating the label, we asked them to import the CSV into the Google Play developer console to create the label. 
Participants were asked to think aloud during both tasks.
We discuss the potential implications of a learning effect due to the two-task within-subjects study design along with other methodological limitations in \autoref{sec:methodological-limitations}.

The study ended with a brief semi-structured interview.
We showed the discrepancies between the two labels and asked the participant to identify errors in either version and explained what caused the error.
Then we asked which tool they preferred and why.
The interview script is in \autoref{sec:interview-script}.

\subsection{Ethics of the Study}
The study was IRB-approved.
Each interview took 1.5 to 2 hours.
Each participant was compensated \$70.
During the interview, we allowed them to only share part of their screen to avoid showing identifiable information and skip any questions they did not feel comfortable answering and reassured them that it would not affect their compensation.
We removed any information that could identify the participants, their apps, and their organizations before publishing the results.

\subsection{Qualitative Analysis Method}

The first author coded the interview transcripts and the screen recordings.
First, the author coded when developers added annotations and modified the XML spec entries.
Then the author coded the verbal responses from the think-aloud process of the main tasks and the post-study interviews using a bottom-up open coding method~\cite{saldana2015coding}.
The other two authors met with the first author weekly to discuss the findings and derive themes.
The coding process was done with the software MAXQDA.
We provide our complete codebook in \autoref{sec:codebook}.

\subsection{Methodological Limitations}
\label{sec:methodological-limitations}

Our study method has some limitations.
First, using Matcha after using the developer console may result in a learning effect: the increased familiarity with the task might have contributed to the improvement in accuracy caused by Matcha. 
However, the process of using the two tools and the questions the developers answer are quite different, which suggests the learning effect may be small.
Our qualitative analysis further delineates how Matcha improved the accuracy.
Second, the identified errors are not exhaustive since the ground truth is not available.
As such, the errors analyzed in the paper should not be interpreted as all possible errors.
Third, our participants developed the app individually or in a small team, so our findings may not apply to developers who work in a big company.
Fourth, our findings be subject to the social desirability bias, namely the participants may be more likely to express a preference for Matcha due to the financial compensation for participation.
Hence, future field research is needed to investigate how developers perceive the tradeoff of time for accuracy in real life.
Fifth, testing Matcha with different apps has inherent limitations as discussed in \autoref{sec:different-project-limitations}.
One potential mitigating approach is to conduct the studies with multiple developers working on the same project.
We decide to leave the exploration of this idea for future research.

\section{Results}

Overall, we found that Matcha helped improve the accuracy of the safety labels.
Both objective and subjective results suggest that Matcha was easy to learn and use.
All participants preferred Matcha over Google's tool.

\subsection{Matcha Improved Label Accuracy}
\label{sec:matcha-improved-label-accuracy}
All participants considered the Matcha version correct when reviewing the discrepancies between the two labels, except for \efour{}, who correctly classified gender as ``other personal information'' in the baseline while then misclassifying it as ``sexual orientation'' using Matcha.
The misclassification errors do not affect counting the data types and purposes and therefore do not affect our quantitative measurements.
All participants except for \eeight{} fixed some errors in their labels with Matcha.
The proportion of participants who have fixed errors in their labels using Matcha was 11/12 (approximately
92\%) with a Wilson confidence interval of (64.6\%, 98.5\%) at the 95\% confidence level~\cite{wilson1927probable, lubin2021statically}.
Matcha helped report 1.8 times as many data types collected or shared by the app (92 vs. 52) and 3.0 times as many purposes for data collection and sharing (212 vs. 70) as compared to the baseline.

\subsection{Types of Errors Fixed by Matcha}
\label{sec:quant-analysis-error}

Our first analysis examines the errors corrected by Matcha in various aspects.
\autoref{tab:fixed-error-data-type-purpose-overview} summarizes the errors.

\begin{table}[]
    \centering
    \caption{Analysis of errors fixed by Matcha. The \textit{Base.} column shows the number of data types and purposes reported using the developer console, and the \textit{Add} and \textit{Cut} columns show what Matcha helped add or remove as compared to the baseline version. Most fixed errors were under-reporting errors (more added than removed) caused by third-party libraries.}
    \resizebox{\columnwidth}{!}{
    \begin{tabular}{p{0.27\linewidth} R{0.1\linewidth} R{0.08\linewidth} R{0.07\linewidth}  R{0.1\linewidth} R{0.08\linewidth} R{0.07\linewidth}}
    \toprule
     & \multicolumn{3}{c}{\textbf{Data Type}} & \multicolumn{3}{c}{\textbf{Data Purpose}} \\
     & Base. & Add & Cut & Base. & Add & Cut\\
     \midrule
    1\textsuperscript{st}-party collect & 21 & 8 & 13 & 34 & 15 & 20\\
    3\textsuperscript{rd}-party collect & 18 & 44 & 11 & 22 & 107 & 15\\
    1\textsuperscript{st}-party share & 1 & 2 & 1 & 2 & 2 & 2\\
    3\textsuperscript{rd}-party share & 12 & 20 & 9 & 12 & 64 & 9\\
    \midrule
    Total & 52 & 74 & 34 & 70 & 188 & 46 \\
    \bottomrule
    \end{tabular}
    }
    \label{tab:fixed-error-data-type-purpose-overview}
\end{table}

\subsubsection{Under-reporting vs. over-reporting}

Matcha fixed many under-reporting errors (77\%), which helped developers report more comprehensive data practices.
\label{sec:quant-analysis-error-matcha-more-comprehensive}
We want to note that the Matcha labels also fixed under-reporting errors in the real-world labels of the six Google Play apps.
Matcha also fixed some over-reporting errors (23\%). 

\subsubsection{First-party vs. third-party}
\label{sec:quant-analysis-error-source}

A significant fraction of errors corrected by Matcha were caused by third-party libraries that automatically collect or share data (78\%) than by first-party code (22\%).
This was mostly due to the Firebase services used for functionality and analytics, as well as other advertising and utility libraries.

\subsubsection{Fixed errors related to different data practices}

More errors fixed by Matcha were related to data collection (70\%) than data sharing (30\%).
However, we want to note that the improvement for data sharing might be more essential, because no data sharing was reported in baseline labels, whereas some data collection was already reported in baseline.
This suggests developers had more severe awareness gaps regarding data sharing.
Furthermore, sharing data with third parties is more sensitive~\cite{chitkara2017does}, which means the Matcha labels can better inform users of the privacy risks.

\subsection{Matcha Helped Tackle Challenges for Creating Accurate Privacy Labels}
\label{sec:qual-eval-error-cause}

We identified four themes in participants' explanations of errors fixed by Matcha.
We found that Matcha helped address common issues that can lead to misunderstanding of data practices and inaccurate privacy nutrition labels.~\cite{li2018coconut, li2022understanding, balebako2014privacy}.

\subsubsection{Help tackle misunderstandings about data safety label (\efour{}, \fone{}, \ftwo{}, \fsix{}, \fthree{}, \fseven{}, \fnine{}, \feleven{})}
\label{sec:qual-error-cause-label-misunderstanding}

Matcha helped fix errors due to misunderstandings of the data safety label taxonomy.
\label{sec:qual-error-cause-overreporting-due-to-term-misunderstanding}
For example, \fsix{} initially thought he should report some data as collected 
while the data was only used on device and therefore did not count as collection per Google's definition.
Matcha explicitly asked whether the data is transmitted off the device, which resolved the problem by relieving the developer from translating low-level behaviors to the label terms.
Interestingly, we observed that developers ignored unfamiliar data types in the baseline task.
For example, \efour{} said \blockquote[]{I did not even think of `other user-generated content.'}
Matcha correct these errors by having them focus on a specific API call and the data types related to the API.

\subsubsection{Help reduce errors related to third-party libraries (\uten{}, \efour{}, \esix{}, \ftwo{}, \fsix{}, \fthree{}, \fnine{})}
\label{sec:qual-error-cause-third-party-misperceptions}

We found Matcha helped correct errors due to misperceptions about third-party libraries.
Some developers did not consider these libraries when creating the label.
For example, \esix{} felt the data collected by Firebase was \blockquote{collected by a different platform} not part of his app.
Some developers were unaware of data collected and shared by libraries.
For example, \efour{} explained that \blockquote{I didn't know that the library (Firebase cloud storage) was doing that on its own behind the scenes.}

Interestingly, Matcha has also helped developers who already had some expectation of the third-party data practices. 
For example, \uten{} said \blockquote{our data is sent to Firebase server, that's why I am selecting these} as he thought out aloud when using the baseline tool.
However, later Matcha revealed that Firebase collected more data then he expected.
\label{sec:in-depth-search-sdk-example}
\fnine{} searched the exact data practices of an advertising library used in his app, but couldn't find the specific guide provided by the library developer, so he referred to their privacy policy instead.
The ambiguous wording of the privacy policy then caused errors in the first label that were later corrected by Matcha.

\subsubsection{Help reduce errors due to forgetfulness (\uten{}, \efour{}, \esix{}, \eeight{}, \ffive{}, \fseven{}, \feleven{})}
\label{sec:qual-error-cause-forgetfulness}

One common source of errors that Matcha helped fix is due to forgetfulness.
They could simply be an oversight -- \blockquote[\esix{}]{it just escaped my mind}, or have deeper reasons.
For example, \eeight{} forgot the use of a third-party library and explained that \blockquote{I didn't actually recall that because I was not in charge of this part.}
\ffive{} forgot he integrated the Admob library a long time ago.
Both errors were caught by Matcha.
Similar to earlier research~\cite{li2022understanding}, we found developers mostly answered questions from memory when using the developer console.
Matcha's systematic review of data practices helped developers find and disclose more data types than they would have otherwise.
For example, although \efour{} checked the Firebase database in the first task, he forgot certain tables and therefore missed certain collected data types.
This was later fixed by Matcha.

\subsubsection{Help reduce errors due to unfamiliar APIs (\uten{}, \efour{})}
\label{sec:qual-error-cause-technical-gap}
Matcha even helped developers learn more about the behavior of unfamiliar APIs.
For example, \efour{} searched for the \verb|LocationManager.getLastKnownLocation| API online when adding annotations for the API, and therefore learned more about the precisions of the location data. 
\uten{} thought the Admob library could not obtain approximate location data because the app did not request location permissions.
However, he later learned from Matcha that the Admob library used the user's IP address to derive the approximate location, which was not controlled by the permission system.

\subsection{Perceived Benefits and Problems of Matcha}
\label{sec:qual-analysis-usability}

All participants preferred Matcha over the baseline due to four main benefits.
We also discuss needs for future improvement.

\subsubsection{Benefit: Improved accuracy (\uten{}, \efour{}, \fone{}, \ffive{}, \fsix{}, \fseven{}, \fnine{})}
The primary benefit of using Matcha was the improved accuracy, which could outweigh the time cost.
As stated by \efour{}, \blockquote{Accuracy wise, I would prefer the tool Matcha...For an app that I'm going to publish on Google Play, I would use Matcha, because like, I emphasize accuracy over efficiency.}

\subsubsection{Benefit: Ease of use (\efour{}, \ftwo{}, \eeight{}, \ffive{}, \fthree{}, \fnine{}, \feleven{})}
\label{sec:qual-benefit-ease-of-use}
Many participants considered Matcha easy to learn and use.
\eeight{} felt the quickfixes for adding annotations were \blockquote{pretty convenient}.
\ffive{} said \blockquote{I thought it might be complicated. But when I started a bit, it becomes easier to use}. 
\efour{} felt it would be easier if he added annotations as he coded the app.

We observed that those who devoted more effort towards providing accurate information in the initial task via the developer console tended to find more value in the ease of use of Matcha. For example, \fnine{} tried to search for the third-party SDKs' data practices in the baseline task (as mentioned in \autoref{sec:in-depth-search-sdk-example}).
He expressed a preference for Matcha because, \blockquote{it is very easy, it saves a lot of time, and plus it is more accurate as compared to the Google Play console, which is very lengthy, and we have to read through all the options and then check the boxes, and we have to consult the documentation.}

\subsubsection{Benefit: Informative tool (\efour{}, \eeight{}, \ftwo{}, \ffive{}, \fsix{}, \fthree{})}
\label{sec:qual-benefit-informative-tool}
Many participants liked Matcha because it helped them learn a lot about their app and the data safety label.
For example, \fsix{} mentioned that, \blockquote{Before using your plugin, I was quite sure that I have submitted all the information that I'm getting, but after using the plugin, I am more knowledge about what's going on in my app.}

\subsubsection{Benefit: Better engagement (\efour{}, \eeight{}, \fseven{})}
\label{sec:qual-benefit-better-engagement}
Some participants liked that Matcha contextualized all the questions around specific code, which better engaged them with this task than the developer console.
\eeight{} explained that: \blockquote{The developer console does have everything written on it, but it's hard to actually relate that to your own code, because it's just a bunch of instructions. While the plugin could remind you of what you have written. For my example, I don't actually remember if I ever imported the WeChat SDK. I really don't remember that. And the console wouldn't actually remind me of anything.}

\subsubsection{Benefit: Better flexibility (\esix{}, \eeight{}, \feleven{})}
\label{sec:qual-benefit-better-flexibility}
Some participants considered the code-based label generation more flexible.
\esix{} mentioned that he preferred the plugin because \blockquote{It gave me the flexibility I needed and made me feel like I was still doing development work. All I had to do was add annotations and it generated labels for me.}
\feleven{} even thought of the benefits of Matcha's annotation-based method for future app updates:
\blockquote{If I update the app to another version, it's easy to change the annotation and create a new CSV file.}

\subsubsection{Problem: Redundancy issues (\efour{}, \eeight{}, \ftwo{}, \fsix{})}
\label{sec:qual-challenge-redundancy}
Developers raised redundancy as a recurrent issue with Matcha, particularly in keyword search results.
Some keywords were too generic and resulted in false positives.
For example, \eeight{} felt that the keyword ``search''  was not effective in detecting the data type ``search history'' because it was often used for unrelated purposes.
Developers also noted the redundancy in the required developer input.
We requested that developers add different annotations when making API calls that collect the same type of data, as they may serve different purposes.
However, \efour{} complained about having to annotate \verb|requestLocationUpdates| after annotating \verb|getLastKnownLocation|, which collect the same data and for the same purpose in his use case.
Future work may consider improving the design with more context-aware suggestions.

\subsection{Efficiency of Matcha}

Our analysis of developers' action traces offers insights into the efficiency and learning curve of Matcha.

\subsubsection{Overall time performance comparison}

It took our participants 30 minutes on average to complete the task using Matcha ($std=15$ minutes), while it only took 9.8 minutes using the developer console ($std=9.3$ minutes).
Although developers were able to complete the task faster using the developer console, they did so at the cost of accuracy.

\subsubsection{Time for adding annotations}

\begin{figure}
\centering
\begin{subfigure}[b]{0.49\linewidth}
    \includegraphics[width=\textwidth]{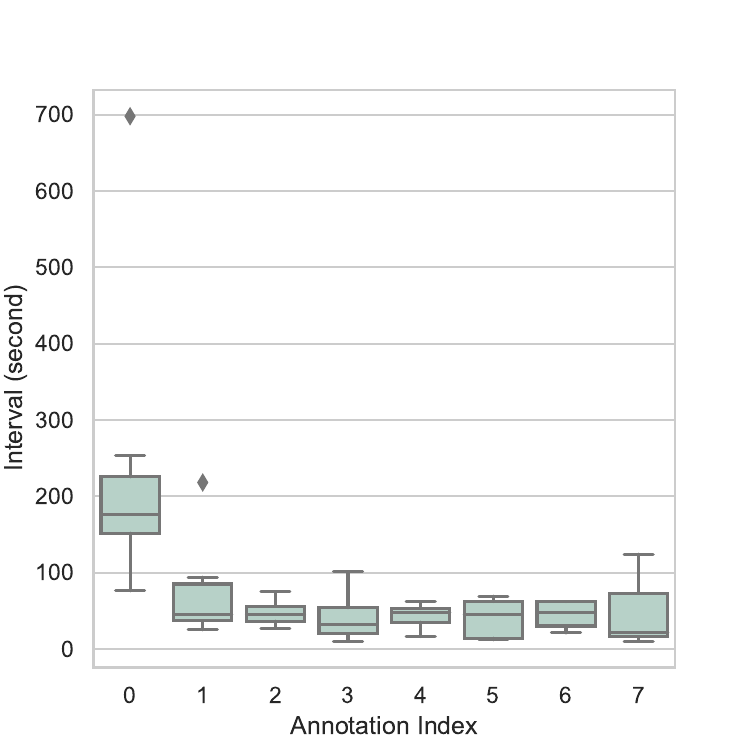}
\caption{Access annotations}
\label{fig:plot_access_annotation_time}
\end{subfigure}
\hfill
\begin{subfigure}[b]{0.49\linewidth}
\includegraphics[width=\textwidth]{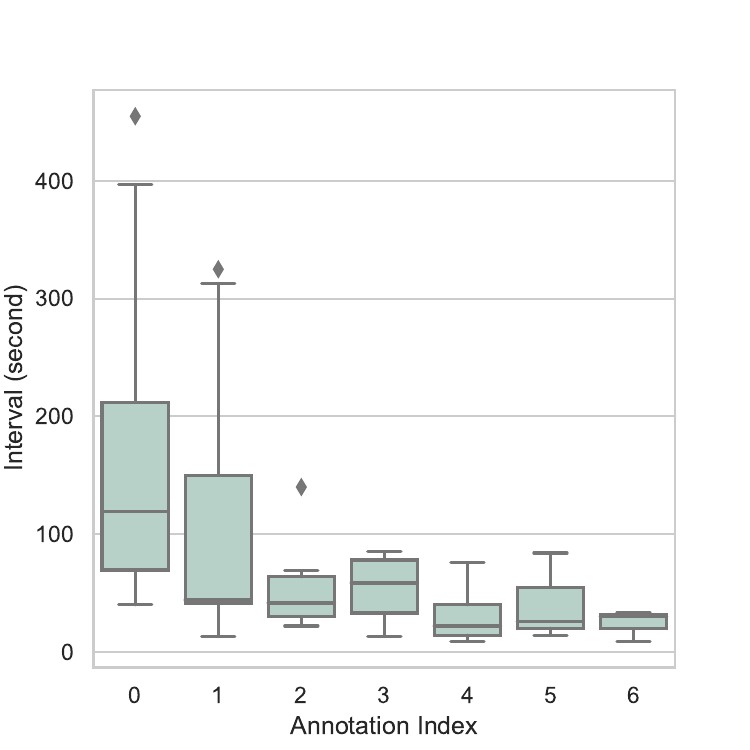}
\caption{Transmission annotations}
\label{fig:plot_transmission_annotation_time}
\end{subfigure}
\caption{Time to add an access and a transmission annotation.
The indices represent the order of added annotations.
For the reliability of the results, we only show the indices with at least three people's data.
The two figures show that it took participants longer to add the first access annotation, while the time drastically decreased after the first two attempts and became stable afterward, suggesting an easy learning curve.}
\label{fig:plot_annotation_time}
\end{figure}

We analyzed the time for adding privacy annotations, which is a novel task for developers.
Each access annotation (\verb|@DataAccess| or \verb|@NotPersonalDataAccess|) took only 1.3 minutes on average to add ($std=1.6$ minutes).
Each transmission annotation (\verb|@DataTransmission| or \verb|@NotPersonalDataTransmission|) also took 1.3 minutes on average ($std=1.6$ minutes).
\label{sec:quant-annotation-learning-curve}
We further show in \autoref{fig:plot_annotation_time} that the time of adding an annotation decreases over time, suggesting our participants' performance increased with practice.

\subsection{Developers' Reactions to Suggestions}
\label{sec:dev-reaction-to-matcha}
Finally, we analyze developers' reactions to the detected API calls and libraries. 
Matcha detected 10 access API calls and 6.3 transmission API calls on average per app, which led to 4.9 access annotations and 5.2 transmission annotations added by participants on average per app.
Note that the number of required annotations are fewer than the detected API calls because multiple API calls can share one annotation.

Among the 59 access annotations, 19 were \texttt{@NotPersonalDataAccess}; among the 62 transmission annotations, 34 were \texttt{@NotPersonalDataTransmission}, which demonstrated the developers' abilities to identify and correct the false positives of Matcha suggestions.
For example, the user password and files provided by the app were the two types of data most commonly labeled as \texttt{@NotPersonalDataAccess} in our study;
and data stored locally on device, network request without user data, and data transmission practices that meet the exemption criteria were the common reasons for \texttt{@NotPersonalDataTransmission}.

Furthermore, \eeight{}, \ffive{}, and \fseven{} each added a \texttt{@DataAccess} for a data type that was not covered by API call detection, showing the benefits of drawing on the developers' knowledge of the app to complement the API-based code analysis results.

\section{Discussion}

\subsection{Developers and Code Analysis: Better Together}

The improved label accuracy (\autoref{sec:matcha-improved-label-accuracy} and \autoref{sec:quant-analysis-error}) shows the efficacy of making developers and code analysis handle the part of work they are most effective at and benefit from each other. 
Meanwhile, using Matcha improved the participants' knowledge of their app's data practices and overcome misunderstandings in data safety labels (\autoref{sec:qual-benefit-informative-tool} and \autoref{sec:qual-error-cause-overreporting-due-to-term-misunderstanding}).
It is useful to keep developers informed and involved during this task because their knowledge can complement and refine the code analysis results (\autoref{sec:dev-reaction-to-matcha}).
At present, developers are largely motivated by platform requirements to consider privacy, and their focus is largely limited to how to satisfy specific requirements~\cite{li2021developers}.
Using Matcha showed a nice side effect to evoke more in-depth learning for privacy (\autoref{sec:qual-error-cause-technical-gap}),
which might motivate and support further improvements in app privacy design.

\subsection{Using Annotations as a Uniform Privacy Language for Developers}

Instead of asking developers to directly describe their apps' data practices using the label terms,
Matcha ask developers to add annotations and edit the XML spec, and let the tool translate them to the label terms (\autoref{fig:annotation-dsl-mapping}).
This not only mitigated errors due to developers' misinterpretations of label terms (\autoref{sec:qual-error-cause-label-misunderstanding}), but also helped developers reflect on their data use (\autoref{sec:qual-benefit-better-engagement}).
Moreover, using annotations allows developers to contribute granular privacy information in a contextualized format, which can potentially be used to automatically generate various types of privacy notices to help relieve developers from the work of upgrading privacy interfaces
~\cite{li2021honeysuckle}.
Hence, the annotations and XML specification may have the potential to become a privacy domain-specific language for developers, bridging the low-level code behavior and the high-level user-facing disclosure.

\subsection{Generalizability of Annotation-based Approach for Creating Privacy Labels}

There are many scenarios beyond Android client apps written in Java that require improved developer support for creating accurate privacy notices.
These include apps developed for other platforms (e.g., iOS, cross-platform web apps) and server-side code.
We analyze the feasibility and cost of applying the annotation-based approach to other contexts in the following.

Matcha focuses on generating Google privacy labels, which follow a different design than the Apple privacy labels.
The key concept in Apple privacy label is also ``data collection''.
The term ``collection'' has a similar definition as in Google's context, which means the annotation design for gathering data collection related information can be easily adapted to the iOS context by coding the exemption rules and special considerations into new pre-defined values for the \verb|collectionAttribute| field.
The two platforms also have unique core concepts in their privacy label design, such as ``data sharing'' for Google, and ``data linked to users'' and ``data used to track users.'' for Apple.
As a result, the \verb|sharingAttribute| field in Matcha annotation design should be correspondingly replaced with \verb|linkingAttribute| and \verb|trackingAttribute|.
Specifically, the former is used to describe whether the data is identifiable or stored with other identifiable data, which will requests users to complement server-side data practices in a similar way as in Matcha.
The latter is related to third-party advertising tracking and is directly associated with the use of the \verb|AppTrackingTransparency| framework.

Then we discuss the technical feasibility of migrating the annotation-based approach to other contexts (e.g., iOS apps, cross-platform web apps.)
We expect privacy annotations to be applicable to other languages, because many other languages also support developers to attach metadata to code in different ways.
For example, C\#, Python, and TypeScript support some level of customization for annotations or decorators similar to Java's annotations, so privacy annotations can be relatively easily migrated to apps written in these languages.
For languages that do not have an official support for custom annotation or decorator yet (e.g., Swift), privacy annotations can be added in an alternative format such as free-form comments, or by extending the language syntax and writing a transpiler to do source-to-source translation.

Transferring privacy annotations to an individual programming language is relatively straightforward.
However, a more sophisticated challenge arises when a real-world software system consists of multiple parts written in different languages.
For example, the frontend website may be coded in TypeScript, while the backend server is coded in Python.
As data may be transmitted across these subsystems, there needs to be a unified format to convert privacy annotations from different languages into intermediate files.
These files can then be aggregated to generate a comprehensive picture of the data practices of the entire system.
Aggregating privacy annotations across systems can further consolidate the annotation design and reduce errors, as some fields that previously needed to be manually filled can now be automatically inferred (e.g., server-side data retention practices, which are currently specified in the data transmission annotation as a collection attribute).

\subsection{Limitations of the Annotation and Developer Tool Approach}

Our work is focused on helping benign developers create accurate privacy labels by tackling their knowledge gaps and addressing common misunderstandings.
However, we want to note that an intrinsic limitation of Matcha is that it can not prevent malicious developers from hiding their data practices from users.
These malicious developers may intentionally aim to steal user data, or simply fear that users will be discouraged from installing the app.
To address this issue, further research is needed to aid external parties (e.g., users, app marketplaces) in auditing the privacy labels.
The auditing task is possible for popular third-party SDKs by cross checking the app's privacy label against the data practices disclosed by third-party SDK developers, who are more likely to be held accountable.

Fully automated auditing for less popular third-party SDKs and first-party data practices can be extremely challenging, because external parties are not able to audit how the data is used after it leaves the device.
Nevertheless, the annotation approach potentially provides external parties with a method to partially vet the privacy labels.
By requiring developers to disclose more intermediate data than is directly provided in Google's label creation form, the intermediate data (e.g., whether certain types of data flow from a source to a sink) can be vetted by dynamic analysis techniques.
Overall, we envision that by designing standardized disclosure formats like annotations requiring developers to provide low-level, unambiguous data practices, we can enhance the auditability of privacy notices beyond existing methods such as privacy policies and privacy labels.

\subsection{Design Implications for Developer Tools for Privacy}

We have shown Matcha successfully helped developers improve the accuracy of their data safety label (\autoref{sec:matcha-improved-label-accuracy} and \autoref{sec:quant-analysis-error}), was easy to learn and use (\autoref{sec:qual-benefit-ease-of-use} and \autoref{sec:quant-annotation-learning-curve}), and was preferred by all study participants (\autoref{sec:qual-analysis-usability}).
Below, we synthesize design recommendations for developer tools for creating standardized privacy notices.

\subsubsection{Contextualize the task around code}
When filling out forms on the developer console, developers rarely checked the code to verify their understanding.
Moreover, developers had trouble systematically reviewing the code on their own.
Matcha suggests that showing questions around the related code can help developers provide more accurate answers and learn more about their app (\autoref{sec:qual-benefit-better-engagement}).

\subsubsection{Provide scaffolding}
Despite the promising benefits of making privacy information part of the code, it is difficult for developers to manually handle the task given the complexity of the required information and the difficulty of handling an unfamiliar syntax (i.e., the annotation).
In Matcha, the use of the quickfix dialog helped solicit valid and accurate input and eased the learning curve.
The dialog allows for more space for presenting the full questions in a structured format and verifies the developer's input before it is submitted.
This method both provides sufficient guidance for novice users and flexibility for expert users. 

\subsubsection{From high-level tasks to low-level questions}
Standardized privacy notices need to lump low-level practices into higher-level categories to improve the clarity of the notice to lay people.
However, as developers often have misperceptions of the standard taxonomy~\cite{li2022understanding}, it is helpful to break down each high-level concept into lower-level questions that probe each aspect of the concept separately.
For example, Matcha helped the developers correct their misunderstandings of ``data collecion'' and ``data sharing'' by separating data accesses and transmissions and asking about special cases and exemption conditions explicitly when the developer added transmission annotations (\autoref{sec:qual-error-cause-label-misunderstanding}).

\subsubsection{Use proactive guidance and actionable suggestions}

One key challenge in creating accurate labels is that developers tend to be overconfident in their answers and unaware of their own errors and knowledge gaps (\autoref{sec:pilot-tests-overconfidence} and \autoref{sec:qual-benefit-informative-tool}).
To better engage the developers in this type of tasks, we used proactive guidance such as the just-in-time tooltips and the errors flagged in code for missing annotations (\autoref{fig:got-it-tooltip-examples}).
We also tried to break down the entire task into smaller actionable steps.
Our suggestions of accesses and transmissions are grounded in these actionable steps to force the developer to interact with them and ponder on them.

\subsubsection{Use precise and specific suggestions}
A fundamental challenge in this type of task is to balance the recall and precision of the suggestions.
Our study showed that the precise and specific API-call based suggestions were better received than more generic keyword-based suggestions (\autoref{sec:qual-challenge-redundancy}).
Although it is still necessary to have something like the keyword-based detection that emphasizes a good recall rate, it would be more effective if the precision is also improved so the correct suggestions are not buried in a large volume of irrelevant suggestions.

\subsection{Future Research Directions}

In this work, we took the first step to design developer tools for creating accurate privacy nutrition labels.
Below we discuss challenges that need to be addressed by future research.

\subsubsection{Managing Third-Party SDK Label Information At Scale}

In the long run, developers may want to obtain support for third-party SDKs that are currently not covered, and receive up-to-date suggestions as these SDKs update their data practices.
However, there are two fundamental barriers: 1) some third-party SDK developers do not disclose their data practices for label creation tasks; 2) even for SDKs that do provide such information, the ad-hoc format of disclosure results in ambiguity and makes it difficult to automatically parse the information. 

We propose two potential directions to tackle these barriers for future research.
One idea, independent of the platform, is to automatically crawl and parse third-party SDK privacy label disclosures, and use a developer-sourcing approach to collectively vet, fix, and release the data as part of an open-source effort.
For example, if a third-party SDK is detected, the tool may first check if there is a public resource associated with the SDK.
If such a resource exists, the tool may retrieve it and compare it with the latest version to identify any changes in content. Then it can categorize these changes using the terminology of privacy label and request the developers to verify them.
If such a resource can not be automatically detected, the tool may request the developer to search for the resource and provide a pointer.
In this situation, third-party SDK developers only need to post and update the guidelines for fulfilling the platform disclosure requirements on their websites, just as they are currently doing.

Another idea relies on the platform to create a standardized format for third-party SDK developers to disclose their data practices, which can then be automatically integrated into the privacy nutrition label creation process.
The closest real-world implementation of this concept is Apple's privacy manifests, introduced in 2023\footnote{\url{https://developer.apple.com/videos/play/wwdc2023/10060/}}.
Apple requires third-party SDK developers to create privacy manifests, allowing downstream app developers to refer to these manifests when creating their privacy labels.
Future research should explore methods to facilitate the creation and verification of accurate, fine-grained third-party disclosures, and a streamlined process to incorporate this information into user-facing disclosures, such as privacy labels.

\subsubsection{Integrating advanced program analysis techniques}

Our proof-of-concept prototype, built with simple code analysis techniques, already achieved substantial improvement.
However, more precise suggestions can help address the redundancy issues mentioned by participants (\autoref{sec:qual-challenge-redundancy}).
One idea is to enhance the keyword search with large programming models such as Codex~\cite{chen2021evaluating}.
In addition, 
future research can also study how to combine dynamic program analysis techniques to provide post-hoc data transmission monitoring and feedback.

\subsubsection{Increasing incentives for improving accuracy}

Matcha helped developers fix many under-reporting errors, but it can also make the apps look more invasive than apps with less accurate labels.
To solve this problem, the platform should take actions to motivate developers to improve accuracy.
For example, if Google provides some level of verification of the data practices and makes the results visible to both the developer and the end users, it can reward developers who honestly disclose their data practices in the privacy label.

\subsubsection{Designing for other real-world challenges}

Our study focused on creating a data safety label for a completed app, while future research should also explore other use scenarios.
For example, future research may examine how developers add annotations while coding, as discussed in the methodological limitations (\autoref{sec:methodological-limitations}) and by our participants (\autoref{sec:qual-benefit-ease-of-use}).
Furthermore, 
future work should study how to support multiple developers or even people in other roles to coordinate changes in data practices and properly encode them in the annotations.

\section{Conclusions}
In this paper, we present Matcha, an IDE plugin that can help developers create an accurate data safety label.
Matcha leverages automated code analysis to offer developers data use suggestions and allows developers control the label with annotations and an XML spec.
In our studies, Matcha helped our participants improve their app's label accuracy.
Matcha was perceived as easy to learn and use, and was preferred by all participants over Google's tool for the benefits of accuracy, better engagement and flexibility, and providing useful information.
We discussed the design implications on developer tools for the creation of standardized privacy notices.

\begin{acks}
This research was supported in part by the National
Science Foundation under Grant No. CNS-1801472, Innovators Network Foundation, and CMU CyLab Seed Funding.
Tianshi Li was supported in part by the CMU CyLab Presidential Fellowship.
The U.S. Government is authorized to reproduce and distribute reprints for Governmental purposes notwithstanding any copyright notation thereon.
The views and conclusions contained herein are those of the authors and should not be interpreted as necessarily representing the official policies or endorsements, either expressed or implied, of the U.S. Government.
We thank the anonymous reviewers for their constructive feedback.
\end{acks}




\bibliographystyle{plainnat}
\bibliography{\jobname}

\begin{thebibliography}{48}
\providecommand{\natexlab}[1]{#1}
\providecommand{\url}[1]{\texttt{#1}}
\expandafter\ifx\csname urlstyle\endcsname\relax
  \providecommand{\doi}[1]{doi: #1}\else
  \providecommand{\doi}{doi: \begingroup \urlstyle{rm}\Url}\fi

\bibitem[Amershi et~al.(2019)Amershi, Weld, Vorvoreanu, Fourney, Nushi, Collisson, Suh, Iqbal, Bennett, Inkpen, et~al.]{amershi2019guidelines}
Saleema Amershi, Dan Weld, Mihaela Vorvoreanu, Adam Fourney, Besmira Nushi, Penny Collisson, Jina Suh, Shamsi Iqbal, Paul~N Bennett, Kori Inkpen, et~al.
\newblock Guidelines for human-ai interaction.
\newblock In \emph{Proceedings of the 2019 chi conference on human factors in computing systems}, pages 1--13, 2019.

\bibitem[Arzt et~al.(2014)Arzt, Rasthofer, Fritz, Bodden, Bartel, Klein, Le~Traon, Octeau, and McDaniel]{arzt2014flowdroid}
Steven Arzt, Siegfried Rasthofer, Christian Fritz, Eric Bodden, Alexandre Bartel, Jacques Klein, Yves Le~Traon, Damien Octeau, and Patrick McDaniel.
\newblock Flowdroid: precise context, flow, field, object-sensitive and lifecycle-aware taint analysis for android apps: precise context, flow, field, object-sensitive and lifecycle-aware taint analysis for android apps.
\newblock \emph{ACM SIGPLAN Notices}, 49\penalty0 (6):\penalty0 259–269, June 2014.
\newblock ISSN 1558-1160.
\newblock \doi{10.1145/2666356.2594299}.
\newblock URL \url{http://dx.doi.org/10.1145/2666356.2594299}.

\bibitem[Balash et~al.(2022)Balash, Ali, Wu, Kanich, and Aviv]{balash2022longitudinal}
David~G Balash, Mir~Masood Ali, Xiaoyuan Wu, Chris Kanich, and Adam~J Aviv.
\newblock Longitudinal analysis of privacy labels in the apple app store.
\newblock \emph{arXiv preprint arXiv:2206.02658}, 2022.

\bibitem[Balebako and Cranor(2014)]{balebako2014improving}
Rebecca Balebako and Lorrie Cranor.
\newblock Improving app privacy: Nudging app developers to protect user privacy.
\newblock \emph{IEEE Security \& Privacy}, 12\penalty0 (4):\penalty0 55–58, July 2014.
\newblock ISSN 1558-4046.
\newblock \doi{10.1109/msp.2014.70}.
\newblock URL \url{http://dx.doi.org/10.1109/msp.2014.70}.

\bibitem[Balebako et~al.(2014{\natexlab{a}})Balebako, Marsh, Lin, Hong, and Faith~Cranor]{balebako2014privacy}
Rebecca Balebako, Abigail Marsh, Jialiu Lin, Jason Hong, and Lorrie Faith~Cranor.
\newblock The privacy and security behaviors of smartphone app developers.
\newblock In \emph{Proceedings 2014 Workshop on Usable Security}, USEC 2014. Internet Society, 2014{\natexlab{a}}.
\newblock \doi{10.14722/usec.2014.23006}.
\newblock URL \url{http://dx.doi.org/10.14722/usec.2014.23006}.

\bibitem[Balebako et~al.(2014{\natexlab{b}})Balebako, Shay, and Faith~Cranor]{balebako2014your}
Rebecca Balebako, Richard Shay, and Lorrie Faith~Cranor.
\newblock Is your inseam a biometric? a case study on the role of usability studies in developing public policy.
\newblock In \emph{Proceedings 2014 Workshop on Usable Security}, USEC 2014. Internet Society, 2014{\natexlab{b}}.
\newblock \doi{10.14722/usec.2014.23039}.
\newblock URL \url{http://dx.doi.org/10.14722/usec.2014.23039}.

\bibitem[Chattopadhyay et~al.(2020)Chattopadhyay, Nelson, Au, Morales, Sanchez, Pandita, and Sarma]{chattopadhyay2020tale}
Souti Chattopadhyay, Nicholas Nelson, Audrey Au, Natalia Morales, Christopher Sanchez, Rahul Pandita, and Anita Sarma.
\newblock A tale from the trenches: cognitive biases and software development: cognitive biases and software development.
\newblock In \emph{Proceedings of the ACM/IEEE 42nd International Conference on Software Engineering}, ICSE ’20. ACM, June 2020.
\newblock \doi{10.1145/3377811.3380330}.
\newblock URL \url{http://dx.doi.org/10.1145/3377811.3380330}.

\bibitem[Chen et~al.(2021)Chen, Tworek, Jun, Yuan, Pinto, Kaplan, Edwards, Burda, Joseph, Brockman, et~al.]{chen2021evaluating}
Mark Chen, Jerry Tworek, Heewoo Jun, Qiming Yuan, Henrique Ponde de~Oliveira Pinto, Jared Kaplan, Harri Edwards, Yuri Burda, Nicholas Joseph, Greg Brockman, et~al.
\newblock Evaluating large language models trained on code.
\newblock \emph{arXiv preprint arXiv:2107.03374}, 2021.

\bibitem[Chitkara et~al.(2017)Chitkara, Gothoskar, Harish, Hong, and Agarwal]{chitkara2017does}
Saksham Chitkara, Nishad Gothoskar, Suhas Harish, Jason~I. Hong, and Yuvraj Agarwal.
\newblock Does this app really need my location?: Context-aware privacy management for smartphones: Context-aware privacy management for smartphones.
\newblock \emph{Proceedings of the ACM on Interactive, Mobile, Wearable and Ubiquitous Technologies}, 1\penalty0 (3):\penalty0 1–22, September 2017.
\newblock ISSN 2474-9567.
\newblock \doi{10.1145/3132029}.
\newblock URL \url{http://dx.doi.org/10.1145/3132029}.

\bibitem[Cooper(2016)]{saldana2015coding}
Robin Cooper.
\newblock Decoding coding via the coding manual for qualitative researchers by johnny saldaña.
\newblock \emph{The Qualitative Report}, October 2016.
\newblock ISSN 1052-0147.
\newblock \doi{10.46743/2160-3715/2009.2856}.
\newblock URL \url{http://dx.doi.org/10.46743/2160-3715/2009.2856}.

\bibitem[Drosos et~al.(2020)Drosos, Barik, Guo, DeLine, and Gulwani]{drosos2020wrex}
Ian Drosos, Titus Barik, Philip~J. Guo, Robert DeLine, and Sumit Gulwani.
\newblock Wrex: A unified programming-by-example interaction for synthesizing readable code for data scientists.
\newblock In \emph{Proceedings of the 2020 CHI Conference on Human Factors in Computing Systems}, CHI ’20. ACM, April 2020.
\newblock \doi{10.1145/3313831.3376442}.
\newblock URL \url{http://dx.doi.org/10.1145/3313831.3376442}.

\bibitem[Emami-Naeini et~al.(2020)Emami-Naeini, Agarwal, Faith~Cranor, and Hibshi]{emami2020ask}
Pardis Emami-Naeini, Yuvraj Agarwal, Lorrie Faith~Cranor, and Hanan Hibshi.
\newblock Ask the experts: What should be on an iot privacy and security label?
\newblock In \emph{2020 IEEE Symposium on Security and Privacy (SP)}. IEEE, May 2020.
\newblock \doi{10.1109/sp40000.2020.00043}.
\newblock URL \url{http://dx.doi.org/10.1109/sp40000.2020.00043}.

\bibitem[Enck et~al.(2014)Enck, Gilbert, Han, Tendulkar, Chun, Cox, Jung, McDaniel, and Sheth]{enck2014taintdroid}
William Enck, Peter Gilbert, Seungyeop Han, Vasant Tendulkar, Byung-Gon Chun, Landon~P. Cox, Jaeyeon Jung, Patrick McDaniel, and Anmol~N. Sheth.
\newblock Taintdroid: An information-flow tracking system for realtime privacy monitoring on smartphones: An information-flow tracking system for realtime privacy monitoring on smartphones.
\newblock \emph{ACM Transactions on Computer Systems}, 32\penalty0 (2):\penalty0 1–29, June 2014.
\newblock ISSN 1557-7333.
\newblock \doi{10.1145/2619091}.
\newblock URL \url{http://dx.doi.org/10.1145/2619091}.

\bibitem[Fowler(2021)]{iPhoneap13:online}
Geoffrey~A. Fowler.
\newblock iphone app privacy labels are a great idea, except when apple lets them deceive - the washington post.
\newblock \url{https://web.archive.org/web/20220630055538/https://www.washingtonpost.com/technology/2021/01/29/apple-privacy-nutrition-label/}, 1 2021.
\newblock (Accessed on 08/27/2022).

\bibitem[Gardner et~al.(2022)Gardner, Feng, Reiman, Lin, Jain, and Sadeh]{gardner2022helping}
Jack Gardner, Yuanyuan Feng, Kayla Reiman, Zhi Lin, Akshath Jain, and Norman Sadeh.
\newblock Helping mobile application developers create accurate privacy labels.
\newblock In \emph{2022 IEEE European Symposium on Security and Privacy Workshops (EuroS\&PW)}. IEEE, June 2022.
\newblock \doi{10.1109/eurospw55150.2022.00028}.
\newblock URL \url{http://dx.doi.org/10.1109/eurospw55150.2022.00028}.

\bibitem[Gordon et~al.(2015)Gordon, Kim, Perkins, Gilham, Nguyen, and Rinard]{gordon2015information}
Michael~I. Gordon, Deokhwan Kim, Jeff Perkins, Limei Gilham, Nguyen Nguyen, and Martin Rinard.
\newblock Information-flow analysis of android applications in droidsafe.
\newblock In \emph{Proceedings 2015 Network and Distributed System Security Symposium}, NDSS 2015. Internet Society, 2015.
\newblock \doi{10.14722/ndss.2015.23089}.
\newblock URL \url{http://dx.doi.org/10.14722/ndss.2015.23089}.

\bibitem[Guo(2021)]{guo2021ten}
Philip Guo.
\newblock Ten million users and ten years later: Python tutor’s design guidelines for building scalable and sustainable research software in academia.
\newblock In \emph{The 34th Annual ACM Symposium on User Interface Software and Technology}, UIST ’21. ACM, October 2021.
\newblock \doi{10.1145/3472749.3474819}.
\newblock URL \url{http://dx.doi.org/10.1145/3472749.3474819}.

\bibitem[Hennink et~al.(2016)Hennink, Kaiser, and Marconi]{hennink2017code}
Monique~M. Hennink, Bonnie~N. Kaiser, and Vincent~C. Marconi.
\newblock Code saturation versus meaning saturation: How many interviews are enough?: How many interviews are enough?
\newblock \emph{Qualitative Health Research}, 27\penalty0 (4):\penalty0 591–608, September 2016.
\newblock ISSN 1552-7557.
\newblock \doi{10.1177/1049732316665344}.
\newblock URL \url{http://dx.doi.org/10.1177/1049732316665344}.

\bibitem[Huang et~al.(2015)Huang, Li, Xiao, Wu, Lu, Zhang, and Jiang]{huang2015supor}
Jianjun Huang, Zhichun Li, Xusheng Xiao, Zhenyu Wu, Kangjie Lu, Xiangyu Zhang, and Guofei Jiang.
\newblock $\{$SUPOR$\}$: Precise and scalable sensitive user input detection for android apps.
\newblock In \emph{24th USENIX Security Symposium (USENIX Security 15)}, pages 977--992, 2015.

\bibitem[Jin et~al.(2018)Jin, Liu, Dodhia, Li, Srivastava, Fredrikson, Agarwal, and Hong]{jin2018they}
Haojian Jin, Minyi Liu, Kevan Dodhia, Yuanchun Li, Gaurav Srivastava, Matthew Fredrikson, Yuvraj Agarwal, and Jason~I. Hong.
\newblock Why are they collecting my data?: Inferring the purposes of network traffic in mobile apps: Inferring the purposes of network traffic in mobile apps.
\newblock \emph{Proceedings of the ACM on Interactive, Mobile, Wearable and Ubiquitous Technologies}, 2\penalty0 (4):\penalty0 1–27, December 2018.
\newblock ISSN 2474-9567.
\newblock \doi{10.1145/3287051}.
\newblock URL \url{http://dx.doi.org/10.1145/3287051}.

\bibitem[Kelley et~al.(2009)Kelley, Bresee, Cranor, and Reeder]{kelley2009}
Patrick~Gage Kelley, Joanna Bresee, Lorrie~Faith Cranor, and Robert~W. Reeder.
\newblock A “nutrition label” for privacy.
\newblock In \emph{Proceedings of the 5th Symposium on Usable Privacy and Security}, SOUPS ’09. ACM, July 2009.
\newblock \doi{10.1145/1572532.1572538}.
\newblock URL \url{http://dx.doi.org/10.1145/1572532.1572538}.

\bibitem[Kelley et~al.(2010)Kelley, Cesca, Bresee, and Cranor]{kelley2010}
Patrick~Gage Kelley, Lucian Cesca, Joanna Bresee, and Lorrie~Faith Cranor.
\newblock Standardizing privacy notices: an online study of the nutrition label approach: an online study of the nutrition label approach.
\newblock In \emph{Proceedings of the SIGCHI Conference on Human Factors in Computing Systems}, CHI ’10. ACM, April 2010.
\newblock \doi{10.1145/1753326.1753561}.
\newblock URL \url{http://dx.doi.org/10.1145/1753326.1753561}.

\bibitem[Kelley et~al.(2013)Kelley, Cranor, and Sadeh]{kelley2013privacy}
Patrick~Gage Kelley, Lorrie~Faith Cranor, and Norman Sadeh.
\newblock Privacy as part of the app decision-making process.
\newblock In \emph{Proceedings of the SIGCHI Conference on Human Factors in Computing Systems}, CHI ’13. ACM, April 2013.
\newblock \doi{10.1145/2470654.2466466}.
\newblock URL \url{http://dx.doi.org/10.1145/2470654.2466466}.

\bibitem[Kollnig et~al.(2022)Kollnig, Shuba, Van~Kleek, Binns, and Shadbolt]{kollnig2022goodbye}
Konrad Kollnig, Anastasia Shuba, Max Van~Kleek, Reuben Binns, and Nigel Shadbolt.
\newblock Goodbye tracking? impact of ios app tracking transparency and privacy labels.
\newblock In \emph{2022 ACM Conference on Fairness, Accountability, and Transparency}, FAccT ’22. ACM, June 2022.
\newblock \doi{10.1145/3531146.3533116}.
\newblock URL \url{http://dx.doi.org/10.1145/3531146.3533116}.

\bibitem[Li et~al.(2014)Li, Bartel, Klein, Traon, Arzt, Rasthofer, Bodden, Octeau, and Mcdaniel]{li2014know}
Li~Li, Alexandre Bartel, Jacques Klein, Yves~Le Traon, Steven Arzt, Siegfried Rasthofer, Eric Bodden, Damien Octeau, and Patrick Mcdaniel.
\newblock I know what leaked in your pocket: uncovering privacy leaks on android apps with static taint analysis.
\newblock \emph{arXiv preprint arXiv:1404.7431}, 2014.

\bibitem[Li et~al.(2018)Li, Agarwal, and Hong]{li2018coconut}
Tianshi Li, Yuvraj Agarwal, and Jason~I. Hong.
\newblock Coconut: An ide plugin for developing privacy-friendly apps: An ide plugin for developing privacy-friendly apps.
\newblock \emph{Proceedings of the ACM on Interactive, Mobile, Wearable and Ubiquitous Technologies}, 2\penalty0 (4):\penalty0 1–35, December 2018.
\newblock ISSN 2474-9567.
\newblock \doi{10.1145/3287056}.
\newblock URL \url{http://dx.doi.org/10.1145/3287056}.

\bibitem[Li et~al.(2021{\natexlab{a}})Li, Louie, Dabbish, and Hong]{li2021developers}
Tianshi Li, Elizabeth Louie, Laura Dabbish, and Jason~I. Hong.
\newblock How developers talk about personal data and what it means for user privacy: A case study of a developer forum on reddit: A case study of a developer forum on reddit.
\newblock \emph{Proceedings of the ACM on Human-Computer Interaction}, 4\penalty0 (CSCW3):\penalty0 1–28, January 2021{\natexlab{a}}.
\newblock ISSN 2573-0142.
\newblock \doi{10.1145/3432919}.
\newblock URL \url{http://dx.doi.org/10.1145/3432919}.

\bibitem[Li et~al.(2021{\natexlab{b}})Li, Neundorfer, Agarwal, and Hong]{li2021honeysuckle}
Tianshi Li, Elijah~B. Neundorfer, Yuvraj Agarwal, and Jason~I. Hong.
\newblock Honeysuckle: Annotation-guided code generation of in-app privacy notices: Annotation-guided code generation of in-app privacy notices.
\newblock \emph{Proceedings of the ACM on Interactive, Mobile, Wearable and Ubiquitous Technologies}, 5\penalty0 (3):\penalty0 1–27, September 2021{\natexlab{b}}.
\newblock ISSN 2474-9567.
\newblock \doi{10.1145/3478097}.
\newblock URL \url{http://dx.doi.org/10.1145/3478097}.

\bibitem[Li et~al.(2022{\natexlab{a}})Li, Reiman, Agarwal, Cranor, and Hong]{li2022understanding}
Tianshi Li, Kayla Reiman, Yuvraj Agarwal, Lorrie~Faith Cranor, and Jason~I. Hong.
\newblock Understanding challenges for developers to create accurate privacy nutrition labels.
\newblock In \emph{CHI Conference on Human Factors in Computing Systems}, CHI ’22. ACM, April 2022{\natexlab{a}}.
\newblock \doi{10.1145/3491102.3502012}.
\newblock URL \url{http://dx.doi.org/10.1145/3491102.3502012}.

\bibitem[Li et~al.(2022{\natexlab{b}})Li, Chen, Li, Agarwal, Cranor, and Hong]{li2022measurement}
Yucheng Li, Deyuan Chen, Tianshi Li, Yuvraj Agarwal, Lorrie~Faith Cranor, and Jason~I. Hong.
\newblock Understanding ios privacy nutrition labels: An exploratory large-scale analysis of app store data.
\newblock In \emph{CHI Conference on Human Factors in Computing Systems Extended Abstracts}, CHI ’22. ACM, April 2022{\natexlab{b}}.
\newblock \doi{10.1145/3491101.3519739}.
\newblock URL \url{http://dx.doi.org/10.1145/3491101.3519739}.

\bibitem[Liu et~al.(2022)Liu, Kuznetsov, Kim, Chang, Kittur, and Myers]{liu2022wigglite}
Michael~Xieyang Liu, Andrew Kuznetsov, Yongsung Kim, Joseph~Chee Chang, Aniket Kittur, and Brad~A. Myers.
\newblock Wigglite: Low-cost information collection and triage.
\newblock In \emph{Proceedings of the 35th Annual ACM Symposium on User Interface Software and Technology}, UIST ’22. ACM, October 2022.
\newblock \doi{10.1145/3526113.3545661}.
\newblock URL \url{http://dx.doi.org/10.1145/3526113.3545661}.

\bibitem[Lubin and Chasins(2021)]{lubin2021statically}
Justin Lubin and Sarah~E. Chasins.
\newblock How statically-typed functional programmers write code.
\newblock \emph{Proceedings of the ACM on Programming Languages}, 5\penalty0 (OOPSLA):\penalty0 1–30, October 2021.
\newblock ISSN 2475-1421.
\newblock \doi{10.1145/3485532}.
\newblock URL \url{http://dx.doi.org/10.1145/3485532}.

\bibitem[McDonald and Cranor(2008)]{mcdonald2008cost}
Aleecia~M McDonald and Lorrie~Faith Cranor.
\newblock The cost of reading privacy policies.
\newblock \emph{I/S: A Journal of Law and Policy for the Information Society}, 4:\penalty0 543, 2008.

\bibitem[Mhaidli et~al.(2019)Mhaidli, Zou, and Schaub]{mhaidli2019we}
Abraham~H Mhaidli, Yixin Zou, and Florian Schaub.
\newblock "we can't live without $\{$Them!$\}$" app developers' adoption of ad networks and their considerations of consumer risks.
\newblock In \emph{Fifteenth Symposium on Usable Privacy and Security (SOUPS 2019)}, pages 225--244, 2019.

\bibitem[Nan et~al.(2015)Nan, Yang, Yang, Zhou, Gu, and Wang]{nan2015uipicker}
Yuhong Nan, Min Yang, Zhemin Yang, Shunfan Zhou, Guofei Gu, and XiaoFeng Wang.
\newblock $\{$UIPicker$\}$:$\{$User-Input$\}$ privacy identification in mobile applications.
\newblock In \emph{24th USENIX Security Symposium (USENIX Security 15)}, pages 993--1008, 2015.

\bibitem[Norman(2013)]{norman2013design}
Don Norman.
\newblock \emph{The design of everyday things: Revised and expanded edition}.
\newblock Basic books, 2013.

\bibitem[Octeau et~al.(2013)Octeau, McDaniel, Jha, Bartel, Bodden, Klein, and Le~Traon]{octeau2013effective}
Damien Octeau, Patrick McDaniel, Somesh Jha, Alexandre Bartel, Eric Bodden, Jacques Klein, and Yves Le~Traon.
\newblock Effective inter-component communication mapping in android with epicc: An essential step towards holistic security analysis.
\newblock In \emph{Proceedings of the 22nd USENIX security symposium}, pages 543--558, 2013.

\bibitem[Schaub et~al.()Schaub, Balebako, Durity, and Cranor]{schaub2015design}
Florian Schaub, Rebecca Balebako, Adam~L. Durity, and Lorrie~Faith Cranor.
\newblock \emph{A Design Space for Effective Privacy Notices*}, page 365–393.
\newblock Cambridge University Press.
\newblock \doi{10.1017/9781316831960.021}.
\newblock URL \url{http://dx.doi.org/10.1017/9781316831960.021}.

\bibitem[Tahaei et~al.(2022)Tahaei, Ramokapane, Li, Hong, and Rashid]{tahaei2022charting}
Mohammad Tahaei, Kopo~M. Ramokapane, Tianshi Li, Jason~I. Hong, and Awais Rashid.
\newblock Charting app developers’ journey through privacy regulation features in ad networks.
\newblock \emph{Proceedings on Privacy Enhancing Technologies}, 2022\penalty0 (3):\penalty0 33–56, July 2022.
\newblock ISSN 2299-0984.
\newblock \doi{10.56553/popets-2022-0061}.
\newblock URL \url{http://dx.doi.org/10.56553/popets-2022-0061}.

\bibitem[Wang et~al.(2022{\natexlab{a}})Wang, Epperson, DeLine, and Drucker]{wang2022diff}
April~Yi Wang, Will Epperson, Robert~A DeLine, and Steven~M. Drucker.
\newblock Diff in the loop: Supporting data comparison in exploratory data analysis.
\newblock In \emph{CHI Conference on Human Factors in Computing Systems}, CHI ’22. ACM, April 2022{\natexlab{a}}.
\newblock \doi{10.1145/3491102.3502123}.
\newblock URL \url{http://dx.doi.org/10.1145/3491102.3502123}.

\bibitem[Wang et~al.(2022{\natexlab{b}})Wang, Wang, Drozdal, Muller, Park, Weisz, Liu, Wu, and Dugan]{wang2022documentation}
April~Yi Wang, Dakuo Wang, Jaimie Drozdal, Michael Muller, Soya Park, Justin~D. Weisz, Xuye Liu, Lingfei Wu, and Casey Dugan.
\newblock Documentation matters: Human-centered ai system to assist data science code documentation in computational notebooks.
\newblock \emph{ACM Transactions on Computer-Human Interaction}, 29\penalty0 (2):\penalty0 1–33, January 2022{\natexlab{b}}.
\newblock ISSN 1557-7325.
\newblock \doi{10.1145/3489465}.
\newblock URL \url{http://dx.doi.org/10.1145/3489465}.

\bibitem[Wang et~al.(2015)Wang, Hong, and Guo]{wang2015using}
Haoyu Wang, Jason Hong, and Yao Guo.
\newblock Using text mining to infer the purpose of permission use in mobile apps.
\newblock In \emph{Proceedings of the 2015 ACM International Joint Conference on Pervasive and Ubiquitous Computing}, UbiComp ’15. ACM, September 2015.
\newblock \doi{10.1145/2750858.2805833}.
\newblock URL \url{http://dx.doi.org/10.1145/2750858.2805833}.

\bibitem[Wilson(1927)]{wilson1927probable}
Edwin~B. Wilson.
\newblock Probable inference, the law of succession, and statistical inference.
\newblock \emph{Journal of the American Statistical Association}, 22\penalty0 (158):\penalty0 209–212, June 1927.
\newblock ISSN 1537-274X.
\newblock \doi{10.1080/01621459.1927.10502953}.
\newblock URL \url{http://dx.doi.org/10.1080/01621459.1927.10502953}.

\bibitem[Xiao et~al.(2022)Xiao, Li, Qin, Guan, Bai, Liao, and Xing]{xiao2022lalaine}
Yue Xiao, Zhengyi Li, Yue Qin, Jiale Guan, Xiaolong Bai, Xiaojing Liao, and Luyi Xing.
\newblock Lalaine: Measuring and characterizing non-compliance of apple privacy labels at scale.
\newblock \emph{arXiv preprint arXiv:2206.06274}, 2022.

\bibitem[Zhang et~al.(2022)Zhang, Feng, Yao, Cranor, and Sadeh]{zhang2022usable}
Shikun Zhang, Yuanyuan Feng, Yaxing Yao, Lorrie~Faith Cranor, and Norman Sadeh.
\newblock How usable are ios app privacy labels?
\newblock \emph{Proceedings on Privacy Enhancing Technologies}, 2022\penalty0 (4):\penalty0 204–228, October 2022.
\newblock ISSN 2299-0984.
\newblock \doi{10.56553/popets-2022-0106}.
\newblock URL \url{http://dx.doi.org/10.56553/popets-2022-0106}.

\bibitem[Zhang et~al.(2020)Zhang, Lowmanstone, Wang, and Glassman]{zhang2020interactive}
Tianyi Zhang, London Lowmanstone, Xinyu Wang, and Elena~L. Glassman.
\newblock Interactive program synthesis by augmented examples.
\newblock In \emph{Proceedings of the 33rd Annual ACM Symposium on User Interface Software and Technology}, UIST ’20. ACM, October 2020.
\newblock \doi{10.1145/3379337.3415900}.
\newblock URL \url{http://dx.doi.org/10.1145/3379337.3415900}.

\bibitem[Zheng et~al.(2022)Zheng, Wang, Wang, and Ma]{zheng2022telling}
Chengbo Zheng, Dakuo Wang, April~Yi Wang, and Xiaojuan Ma.
\newblock Telling stories from computational notebooks: Ai-assisted presentation slides creation for presenting data science work.
\newblock In \emph{CHI Conference on Human Factors in Computing Systems}, CHI ’22. ACM, April 2022.
\newblock \doi{10.1145/3491102.3517615}.
\newblock URL \url{http://dx.doi.org/10.1145/3491102.3517615}.

\bibitem[Zimmeck et~al.(2021)Zimmeck, Goldstein, and Baraka]{zimmeck2021privacyflash}
Sebastian Zimmeck, Rafael Goldstein, and David Baraka.
\newblock Privacyflash pro: Automating privacy policy generation for mobile apps.
\newblock In \emph{Proceedings 2021 Network and Distributed System Security Symposium}, NDSS 2021. Internet Society, 2021.
\newblock \doi{10.14722/ndss.2021.24100}.
\newblock URL \url{http://dx.doi.org/10.14722/ndss.2021.24100}.

\end{thebibliography}

\appendix
\pagebreak

\section{Annotation Design Details}
\label{sec:annotation-design-details}

\autoref{tab:collection-attribute-questions} and \autoref{tab:sharing-attribute-questions} summarize the design details of the \verb|@DataTransmission| annotation.

\begin{table*}[]
    \centering
        \caption{The \texttt{collectionAttribute} field of the \texttt{@DataTransmission} annotation encodes the data collection information as a list of predefined attribute values. This table shows the groups of attributes that need to be completed in this field, as well as the corresponding collection questions and the exempt conditions of collection defined by Google.}
\begin{tabular}{p{0.2\linewidth} p{0.3\linewidth} p{0.5\linewidth}}
    \toprule
    Attribute name & Values & Original questions / Exempt conditions \\
    \midrule
    \texttt{TransmittedOffDevice} & True or False & Is this data collected, shared, or both? \\
        \midrule

    \texttt{NotStoredInBackend} & True or False & Is this data processed ephemerally? \\
        \midrule

    \texttt{EncryptedInTransit} & True or False & Is all of the user data collected by your app encrypted in transit? \\
        \midrule

    \texttt{OptionalCollection} & True or False & Is this data required for your app, or can users choose whether it's collected? \\
        \midrule

    \texttt{UserToUserEncryption} & True or False & User data that is sent off device, but that is unreadable by you or anyone other than the sender and recipient as a result of end-to-end encryption does not need to be disclosed. \\
    \midrule
    \texttt{CollectedFor} & Seven options: App functionality; Analytics;  Developer communications; Advertising or marketing;  Fraud prevention, Security and compliance; Personalization; Account Management & Why is this user data collected? Select all that apply. \\
    \bottomrule
\end{tabular}
    \label{tab:collection-attribute-questions}
\end{table*}

\begin{center}
        \tablecaption{The \texttt{sharingAttribute} field of the \texttt{@DataTransmission} annotation encodes the data sharing information as a list of predefined attribute values. This table shows the groups of attributes that need to be completed in this field, as well as the corresponding sharing questions and the exempt conditions of sharing defined by Google.}

\begin{supertabular}{p{0.33\linewidth} p{0.27\linewidth} p{0.4\linewidth}}
    \toprule
    Attribute name & Values & Original questions / Exempt conditions \\
    \midrule
    \texttt{SharedWithThirdParty} & True or False & Is this data collected, shared, or both? \\
        \midrule
    \texttt{OnlySharedWithServiceProviders} & True or False & Sharing is exempt if transferring user data to a “service provider” that processes it on behalf of the developer. \\
        \midrule
    \texttt{OnlySharedForLegalPurposes} & True or False & Sharing is exempt if transferring user data for specific legal purposes, such as in response to a legal obligation or government requests. \\
        \midrule
    \texttt{OnlyInitiatedByUser} & True or False & Sharing is exempt if transferring user data to a third party based on a specific user-initiated action, where the user reasonably expects the data to be shared. \\
        \midrule
    \texttt{OnlyAfterGettingUserConsent} & True or False & Sharing is exempt if transferring user data to a third party based on a prominent in-app disclosure and consent that meets the requirements described in our User Data policy. \\
        \midrule
    \texttt{OnlyTransferringAnonymousData} & True or False & Sharing is exempt if transferring user data that has been fully anonymized so that it can no longer be associated with an individual user. \\
        \midrule
    \texttt{SharedFor} & Seven options: App functionality; Analytics;  Developer communications; Advertising or marketing;  Fraud prevention, Security and compliance; Personalization; Account Management & Why is this user data shared? Select all that apply. \\
    \bottomrule
\end{supertabular}
    \label{tab:sharing-attribute-questions} \\
\end{center}

\section{Matcha Implementation Details}
\label{sec:implementation-details}
This section summarizes details about the implementation of the Matcha IDE plugin.
\autoref{tab:keyword-list-permission-based} and \autoref{tab:keyword-list-others} presents the keywords used in Matcha to facilitate the detection of sensitive data access code.
\autoref{tab:third-party-sdk-list} presents all the SDKs that Matcha can detect and automatically generate the safety label for based on the SDK's open documentation about its data practices.

\begin{center}
    \tablecaption{Matcha keyword list (based on definitions and the keywords extracted from open-sourced projects that contain sensitive API calls). Matcha uses keyword search to complement the API-based detection of code that accesses sensitive user data.}
    \begin{supertabular}{p{0.2\linewidth} p{0.2\linewidth} p{0.55\linewidth}}
    \toprule
    Category & Data Type & Keywords \\
    \midrule
    \multirow{9}{3cm}{Personal Info} & Name & name \\
         & Email Address & email\\
         & User ID & uid, user id\\
         & Address & home address, city, country, zip code\\
         & Phone Number & phone, default dialer \\
         & Race and Ethnicity & race, ethnicity, african, indian, asian\\
         & Political or Religious Beliefs & political, religious\\
         & Sexual Orientation & sexual orientation, gay, lesbian, transgender, bisexual, queer \\
         & Other Personal Info & birth, nationality, gender, male, female, non-binary, veteran \\
    \multirow{4}{3cm}{Financial Info} & User Payment Info & credit card, billing, cvv,
                routing number, account number, bank\\
        & Purchase History & purchase\\
        & Credit Score & credit score\\
        & Other Financial Info & salary, debt\\
    Calendar & Calendar Events & calendar, attendee\\
    \multirow{2}{3cm}{Photos and Videos} & Photos & photo, barcode, image, picture, media\\
    & Videos & video, recording, media\\
    Contacts & Contacts & contact, call history,
                interaction duration\\
    \multirow{2}{3cm}{Location} & Approximate Location & location,
                city, country, ip address\\
    & Precise Location & location,
                latitude, longitude\\
    \multirow{2}{3cm}{Health and Fitness} & Health Info & health, medical, medicine, symptom, disease, doctor, physician, sleep, wellness, therapist, emergency, emergencies,
                period, pregnancy \\
    & Fitness Info & fitness, exercise,
                workout, sport, diet, nutrition\\
    \multirow{3}{3cm}{Messages} & Emails & email, sender, recipient, subject \\
    & Sms or Mms & message, sms, mms, sender, recipient, subject\\
    & In-App Messages & message, chat, reply, replies, comment, sender, recipient, subject\\
    Device or Other IDs & Device or Other IDs & mac address, widevine, device id, instance id, app id, advertising id, fingerprint, user agent, unique id, token, AdvertisingIdClient\\
    Files and Docs & Files and Docs & file, document, backup, restore, download, storage, media\\
    \multirow{3}{3cm}{Audio Files} & Voice or Sound Recordings & voice, sound, recording\\
    & Music Files & music, song\\
    & Other User Audio Files & \\
    \multirow{5}{3cm}{App Activity} & App Interactions & selected, visit number, view number, getItemAtPosition, getItemIdAtPosition, AccessibilityService, TextService, Instrumentation, shortcut\\
    & Installed Apps & installed app \\
    & In-App Search History & search\\
    & Other User-Generated Content & bios, note, response\\
    & Other User Activities & gameplay, dialog option\\
    Web Browsing & Web Browsing History & browser, cookie, browser cache, browsing cache, search, web view\\
    \multirow{3}{3cm}{App Info and Performance} & Crash Logs & crash, stack trace\\
    & Diagnostics & ActivityManager, ApplicationErrorReport, ApplicationExitInfo, BatteryManager, Benchmark, Debug, HealthStats, Macrobenchmark, PowerManager, StrictMode, battery, loading time, latency, frame rate, diagnostics\\
    & Other App Performance Data & performance\\
    \bottomrule
\end{supertabular}       
\label{tab:keyword-list-others}
\end{center}



\begin{center}
    \tablecaption{Matcha keyword list (based on permissions). Matcha uses keyword search to complement the API-based detection of code that accesses sensitive user data.}
\begin{supertabular}{p{0.2\linewidth} p{0.2\linewidth} p{0.55\linewidth}}
    \toprule
    Category & Data Type & Keywords \\
    \midrule
    \multirow{9}{3cm}{Personal Info} & Name & BIND\_AUTOFILL\_SERVICE, GET\_ACCOUNTS \\
         & Email Address & BIND\_AUTOFILL\_SERVICE, GET\_ACCOUNTS \\
         & User ID & BIND\_AUTOFILL\_SERVICE, GET\_ACCOUNTS\\
         & Address & BIND\_AUTOFILL\_SERVICE, GET\_ACCOUNTS\\
         & Phone Number & BIND\_AUTOFILL\_SERVICE, GET\_ACCOUNTS, READ\_CALL\_LOG, READ\_PHONE\_NUMBERS, READ\_PHONE\_STATE, READ\_SMS\\
         & Race and Ethnicity & BIND\_AUTOFILL\_SERVICE, GET\_ACCOUNTS\\
         & Political or Religious Beliefs & BIND\_AUTOFILL\_SERVICE,
                GET\_ACCOUNTS\\
         & Sexual Orientation & BIND\_AUTOFILL\_SERVICE, GET\_ACCOUNTS\\
         & Other Personal Info & BIND\_AUTOFILL\_SERVICE,
                GET\_ACCOUNTS\\
    \multirow{4}{3cm}{Financial Info} & User Payment Info & BIND\_AUTOFILL\_SERVICE\\
        & Purchase History & \\
        & Credit Score & \\
        & Other Financial Info & \\
    Calendar & Calendar Events & READ\_CALENDAR, WRITE\_CALENDAR \\
    \multirow{2}{3cm}{Photos and Videos} & Photos & READ\_EXTERNAL\_STORAGE, WRITE\_EXTERNAL\_STORAGE \\
    & Videos & READ\_EXTERNAL\_STORAGE, WRITE\_EXTERNAL\_STORAGE\\
    Contacts & Contacts & ACCEPT\_HANDOVER, ADD\_VOICEMAIL, ANSWER\_PHONE\_CALLS,
                CALL\_PHONE, PROCESS\_OUTGOING\_CALLS, READ\_CALL\_LOG, READ\_CONTACTS, READ\_PHONE\_NUMBERS,
                READ\_PHONE\_STATE, READ\_SMS, RECEIVE\_MMS, RECEIVE\_SMS, RECEIVE\_WAP\_PUSH, SEND\_SMS,
                WRITE\_CONTACTS\\
    \multirow{2}{3cm}{Location} & Approximate Location & ACCESS\_COARSE\_LOCATION, ACCESS\_MEDIA\_LOCATION\\
    & Precise Location & ACCESS\_FINE\_LOCATION, ACCESS\_MEDIA\_LOCATION\\
    \multirow{2}{3cm}{Health and Fitness} & Health Info & ACTIVITY\_RECOGNITION, BODY\_SENSORS\\
    & Fitness Info & ACTIVITY\_RECOGNITION, BODY\_SENSORS\\
    \multirow{3}{3cm}{Messages} & Emails & \\
    & Sms or Mms & READ\_SMS, RECEIVE\_MMS, RECEIVE\_SMS,
                RECEIVE\_WAP\_PUSH, SEND\_SMS, WRITE\_SMS\\
    & In-App Messages & \\
    Device or Other IDs & Device or Other IDs & AD\_ID, READ\_PRIVILEGED\_PHONE\_STATE\\
    Files and Docs & Files and Docs & READ\_EXTERNAL\_STORAGE,
                WRITE\_EXTERNAL\_STORAGE, MANAGE\_EXTERNAL\_STORAGE \\
    \multirow{3}{3cm}{Audio Files} & Voice or Sound Recordings & CAPTURE\_AUDIO\_OUTPUT, RECORD\_AUDIO,
                READ\_EXTERNAL\_STORAGE, WRITE\_EXTERNAL\_STORAGE\\
    & Music Files & READ\_EXTERNAL\_STORAGE, WRITE\_EXTERNAL\_STORAGE\\
    & Other User Audio Files & CAPTURE\_AUDIO\_OUTPUT, RECORD\_AUDIO,
                READ\_EXTERNAL\_STORAGE, WRITE\_EXTERNAL\_STORAGE\\
    \multirow{5}{3cm}{App Activity} & App Interactions & QUERY\_ALL\_PACKAGES\\
    & Installed Apps & \\
    & In-App Search History & \\
    & Other User-Generated Content & \\
    & Other User Activities & \\
    Web Browsing & Web Browsing History & \\
    \multirow{3}{3cm}{App Info and Performance} & Crash Logs & \\
    & Diagnostics & BATTERY\_STATS\\
    & Other App Performance Data & \\
    \bottomrule
\end{supertabular}
    \label{tab:keyword-list-permission-based}
\end{center}

\begin{center}
    \topcaption{Matcha third-party SDK list. Matcha can automatically detect 58 third-party SDKs and automatically fill out the data collection and sharing practices based on the SDK's documentation. The list is primarily curated based on the Google Play SDK Index and also contains a few SDKs developed by Google which also provided such type of documentation.}

    \begin{supertabular}{p{0.25\linewidth} p{0.75\linewidth}}
    \toprule
    SDK Names & Maven ID Matching Pattern \\
    \midrule
     AdMob    &  .*com.google.android.gms:play-services-ads.*|.*com.google.android.gms:play-services-ads-lite.*\\
     Ironsource & .*com.ironsource.sdk:mediationsdk.*\\
     Vungle & .*com.vungle:publisher-sdk-android.* \\
     AppsFlyer & .*com.appsflyer:af-android-sdk.* \\
     Adjust & .*com.adjust.sdk:adjust-android.* |.*com.android.installreferrer:installreferrer.* |.*com.adjust.sdk:adjust-android-webbridge.* \\
     Chartboost & .*com.chartboost:chartboost-sdk.* \\
     Tapjoy & .*com.tapjoy:tapjoy-android-sdk.* \\
     Google Play Games Services & .*com.google.android.gms:play-services-games.* \\
     Firebase Authentication & .*com.google.firebase:firebase-auth.*|.*com.google.firebase:firebase-auth-ktx.* \\
     Firebase App Check & .*com.google.firebase:firebase-appcheck.*|.*com.google.firebase:firebase-appcheck-debug.*|.*com.google.firebase:firebase-appcheck-safetynet.*
     |.*com.google.firebase:firebase-appcheck-playintegrity.* \\
     Firebase Cloud Firestore & .*com.google.firebase:firebase-firestore.*| .*com.google.firebase:firebase-firestore-ktx.* \\
    Cloud Functions for Firebase & .*com.google.firebase:firebase-functions.*|.*com.google.firebase:firebase-functions-ktx.* \\
    Firebase Cloud Messaging & .*com.google.firebase:firebase-messaging.*|.*com.google.firebase:firebase-messaging-ktx.* \\
    Cloud Storage for Firebase &
    .*com.google.firebase:firebase-storage.*|.*com.google.firebase:firebase-storage-ktx.* \\
    Crashlytics & .*com.google.firebase:firebase-crashlytics.*|.*com.google.firebase:firebase-crashlytics-ktx.*|.*com.google.firebase:firebase-crashlytics-ndk.* \\
    Dynamic Links & .*com.google.firebase:firebase-dynamic-links.*|.*com.google.firebase:firebase-dynamic-links-ktx.* \\
    Google Analytics & .*com.google.firebase:firebase-analytics.*|.*com.google.firebase:firebase-analytics-ktx.* \\
    Firebase In-App Messaging & .*com.google.firebase:firebase-inappmessaging.*|.*com.google.firebase:firebase-inappmessaging-display.*|.*com.google.firebase:firebase-inappmessaging-ktx.*|.*com.google.firebase:firebase-inappmessaging-display-ktx.* \\
    Firebase Installations & .*com.google.firebase:firebase-installations.*|.*com.google.firebase:firebase-installations-ktx.* \\
    Firebase ML model downloader & .*com.google.firebase:firebase-ml-modeldownloader.*|.*com.google.firebase:firebase-ml-modeldownloader-ktx.* \\
    Performance Monitoring & .*com.google.firebase:firebase-perf.*|.*com.google.firebase:firebase-perf-ktx.* \\
    Realtime Database & .*com.google.firebase:firebase-database.*|.*com.google.firebase:firebase-database-ktx.* \\
    Remote Config & .*com.google.firebase:firebase-config.*|.*com.google.firebase:firebase-config-ktx.* \\
    RevenueCat & .*com.revenuecat.purchases:purchases.*
    |.*com.revenuecat.purchases:purchases-store-amazon.* \\
    User Messaging Platform SDK & .*com.google.android.ump:user-messaging-platform.* \\
    reCAPTCHA Enterprise & .*com.google.android.gms:play-services-recaptcha.* \\
    ARCore & .*com.google.ar:core:.* \\
    ML Kit & .*com.google.android.gms:play-services-mlkit-barcode-scanning.*|.*com.google.android.gms:play-services-mlkit-face-detection.*|.*com.google.android.gms:play-services-mlkit-image-labeling.*|.*com.google.android.gms:play-services-mlkit-image-labeling-custom.*|.*com.google.android.gms:play-services-mlkit-language-id.*|.*com.google.android.gms:play-services-mlkit-text-recognition.*|.*com.google.android.gms:play-services-code-scanner.*|\\
    &    .*com.google.mlkit:barcode-scanning.*|.*com.google.mlkit:camera.*
    |.*com.google.mlkit:digital-ink-recognition.*|.*com.google.mlkit:entity-extraction.*|.*com.google.mlkit:face-detection.*|.*com.google.mlkit:image-labeling.*|.*com.google.mlkit:image-labeling-custom.*|.*com.google.mlkit:language-id.*|.*com.google.mlkit:linkfirebase.*
    |.*com.google.mlkit:object-detection.*|.*com.google.mlkit:object-detection-custom.*|.*com.google.mlkit:playstore-dynamic-feature-support.*|.*com.google.mlkit:pose-detection.*|.*com.google.mlkit:pose-detection-accurate.*
    |.*com.google.mlkit:segmentation-selfie.*|.*com.google.mlkit:smart-reply.*|.*com.google.mlkit:text-recognition.*|.*com.google.mlkit:text-recognition-chinese.*|.*com.google.mlkit:text-recognition-devanagari.*|.*com.google.mlkit:text-recognition-japanese.*|.*com.google.mlkit:text-recognition-korean.*|.*com.google.mlkit:translate.* \\
    Google Cast (cast-tv) & .*com.google.android.gms:play-services-cast-tv.* \\
    Google Maps & .*com.google.android.gms:play-services-maps.* \\
    Google Pay - Wallet SDK & .*com.google.android.gms:play-services-wallet.* \\
    Google Pay - TapandPay SDK & .*com.google.android.gms:play-services-tapandpay.* \\
    SafetyNet & .*com.google.android.gms:play-services-safetynet.* \\
    Google Play Integrity & .*com.google.android.play:integrity.* \\
    Snowplow Android Tracker & .*com.snowplowanalytics:snowplow-android-tracker.*\\
    Kochava & .*com.kochava.base:tracker.* \\
    Airship SDK & .*com.urbanairship.android:urbanairship-fcm.*|.*com.urbanairship.android:urbanairship-hms.*|.*com.urbanairship.android:urbanairship-message-center.*
    |.*com.urbanairship.android:urbanairship-adm.*
    |.*com.urbanairship.android:urbanairship-preference-center.*
    |.*com.urbanairship.android:urbanairship-automation.* \\
    Appodeal SDK for Android & .*com.appodeal.ads:sdk.* \\
    Apptentive & .*com.apptentive:apptentive-android.* \\
    Branch & .*io.branch.sdk.android:library.* \\
    Braze Android SDK & .*com.appboy:android-sdk-ui.* \\
    Bugsnag & .*com.bugsnag:bugsnag-android.* \\
    CleverTap Android SDK & .*com.clevertap.android:clevertap-android-sdk.* \\
    Fyber Marketplace SDK & .*com.fyber:marketplace-sdk.* \\
    HyprMX & .*com.hyprmx.android:HyprMX-SDK.* \\
    Instabug & .*com.instabug.library:instabug.* \\
    Interactive Media Ads (IMA) SDK & .*com.google.ads.interactivemedia.v3:interactivemedia.* \\
    MoEngage Android SDK & .*com.moengage:moe-android-sdk.* \\
    Ogury SDK & .*co.ogury:ogury-sdk.* \\
    Pangle Ad SDK & .*com.pangle.global:ads-sdk.* \\
    Pollfish & .*com.pollfish:pollfish-googleplay.* \\
    PubMatic OpenWrap SDK & .*com.pubmatic.sdk:openwrap.*\\
    Singular SDK & .*com.singular.sdk:singular\_sdk.* \\
    Smaato NextGen SDK & .*com.smaato.android.sdk:smaato-sdk.*|.*com.smaato.android.sdk:smaato-sdk-rewarded-ads.*|.*com.smaato.android.sdk:smaato-sdk-banner.*|.*com.smaato.android.sdk:smaato-sdk-interstitial.* \\
    Start.io (Formerly StartApp) & .*com.startapp:inapp-sdk.* \\
    Taboola SDK & .*com.taboola:android-sdk.* \\
    Verve Group HyBid SDK (formerly PubNative) & .*net.pubnative:hybid.sdk.* \\
    \bottomrule
\end{supertabular}
    \label{tab:third-party-sdk-list}
\end{center}

\section{Pre-Study Survey}
\label{sec:pre-study-survey}

Thank you for agreeing to participate in this CMU study on creating the Google Play data safety section. We look forward to interviewing you.
 
For this interview study, we will ask you to create the data safety section for one Android app that we selected from your recent Android apps. The selected app has been sent to you. If you are not sure which app to report on, please message us to ask.
 
In this pre-study survey, we would like to ask a few questions about you and the selected app. At the end of the survey, you will see a scheduling link where you can make a booking for our interview.

\begin{description}
    \item[(1)] What is your participant ID for this study? (The ID was sent to you.)
    \item[(2)] What is the Google Play link to the app that you will report on? (The selected app was sent to you. If your app is not on Google Play, just provide the name of the app that you entered in the screening survey.)
    \item[(3)] Please confirm that your app is mainly developed in Java (not other languages/frameworks such as Kotlin, Unity, Flutter, Cordova etc.)
    \begin{itemize}
        \item Yes, my app is mainly developed in Java
        \item No, my app is developed in other languages/frameworks
    \end{itemize}
    \item[(4)] Which option best describes this Android app?
    \begin{itemize}
        \item Commercial project
        \item Research project
        \item Course project
        \item Hobby Project
        \item Other [Free form response expected]
    \end{itemize}
    \item[(5)] (If the Q4 answer is commercial project) How many employees work in the company that developed this app?
    \begin{itemize}
        \item 1-4
        \item 5-9
        \item 10-19
        \item 20-49
        \item 50-99
        \item 100-249
        \item 250-499
        \item 500-999
        \item 1,000 or more
    \end{itemize}
    \item[(6)] Is this an individual-developed app or a group-developed app?
    \begin{itemize}
        \item individual
        \item group
    \end{itemize}
    \item[(7)] (if the Q6 answer is individual) How many people participated in the development of this app (including app design, mobile app, and server side development) ?
    \begin{itemize}
        \item 2-5
        \item 6-10
        \item 11-20
        \item more than 20
    \end{itemize}
    \item[(8)] Which of these roles describe your job for developing this app? (Please select all that apply)
    \begin{itemize}
        \item Mobile App Developer
        \item Backend Developer
        \item Data Scientist and Analyst
        \item Designer
        \item Project Manager
        \item Security Engineer
        \item Privacy Engineer
        \item Quality Assurance Analyst
        \item Other roles (please specify) [Free form response expected]
    \end{itemize}
    \item[(9)] Note that during the study, you will try out an Android Studio plugin and use it to generate the data safety label for your selected app. Therefore, please make sure to install Android Studio and have your app's source code readily available on the machine you use for the interview. Since our plugin may guide you to add annotations in your app's code, we recommend you to either create a copy of your app or commit all changes before the study. We need to collect the data safety labels you created during the study solely for research purposes, and we will not collect other data about your app. Feel free to remove the annotations and uninstall the plugin after the study. I have read and understood the requirements above and still want to participate in this study. If you have any concerns, please contact me before submitting the survey.
    \begin{itemize}
        \item Yes
        \item No
    \end{itemize}
    \item[(10)] What is the version of your Android Studio?
    \item[(11)] Are you a professional Software Developer, i.e. software development is the major component of your job? 
    \begin{itemize}
        \item Yes
        \item No
    \end{itemize}
    \item[(12)] Did you major in computer science or related fields in school?
    \begin{itemize}
        \item Yes
        \item No
    \end{itemize}
    \item[(13)] What is your gender?
    \begin{itemize}
        \item Man
        \item Woman
        \item Non-binary/third gender
        \item Prefer not to answer
    \end{itemize}
    \item[(14)] What is your age group?
    \begin{itemize}
        \item 18-24
        \item 25-34
        \item 35-44
        \item 45-54
        \item 55-64
        \item 65+
        \item Prefer not to answer
    \end{itemize}
    \item[(15)] In which country do you currently reside?
\end{description}

\section{Interview Script}
\label{sec:interview-script}

\subsection{Introduction}
Thanks for agreeing to participate in our study. First, I need to read our standard introduction, as required by our study protocol.

Our group at CMU has been doing research for many years on tools for developers. We are currently working on a research project about the Google Play safety labels, which is a new feature of the Google Play store that shows details of Android apps to end users. Android developers are now required to provide the privacy details for their apps by answering certain questions about data collection and sharing. The general goal of our research is to learn about how Android developers accomplish this task to help us improve a developer tool we design and build to streamline this task.

We understand that you have developed an app named [the app name].  We would like to have you complete the task of creating a safety label for the selected app using different methods. We expect the entire study session to take approximately 90 minutes, though timing may vary depending on the complexity of the app. The study involves two tasks. In the first task, we will ask you to create the label on the Google Play developer console. Then we will ask you to create the label again using a different method. Finally, we will ask some follow-up questions regarding the labels you created during the study, how you perceive certain concepts, and whether you encountered any difficulty during the process. Since we want to observe how you completed this task, we would like you to share your screen during the interview. We need to record both the audio and the screen during the entire interview solely for analysis purposes. We will use Zoom to make the recordings. Only researchers in our group working on this project will have access to the recordings. The interviews will be transcribed automatically by Zoom and we may include parts of the transcripts in our research papers that do not identify you, your app, or your organization.

In the second task, we would like you to try out an Android Studio plugin developed by our lab. We would like you to install the plugin on your Android Studio and then open the source code of the selected app in this Android Studio. The plugin will guide you to create a csv file that you can import into the Google Play developer console to complete the safety label requirement. We will ask you to send us the csv file for analysis purposes. The plugin will not collect any other information about your app and you can either choose to keep it installed or remove it after the study.

Do you have the latest version of Android Studio IDE installed? Some features of the plugin may not work well if you’re not using the latest version of Android Studio. [Proceed after getting their affirmative answer]

Do you have the source code prepared on this machine? Is it OK to install the plugin on your Android Studio? [Proceed after getting their affirmative answer]

And in the second task/later part of the study, the plugin may potentially guide you to make some slight modifications to your app, such as adding annotations and adding a configuration file. We highly recommend you to make a copy of your source code or commit all the previous changes before the study. Is this OK with you? [Proceed after getting their affirmative answer]

Your participation is entirely voluntary and you may quit the study at any time. If you don’t feel comfortable answering a question, feel free to skip it and it will not affect your compensation. You must be 18 or older to participate in this study. You will be compensated \$70 for participating. The interview will be conducted remotely through the computer. Since the interview will be recorded, it is important that you be in a private room, and not in an open-space cubicle, for example. These recordings may be stored on protected computers at CMU and on Zoom, with transcripts potentially edited using a service called Otter. There are no expected risks or benefits to you for participating, beyond the benefits of helping improve the understanding of privacy labels in general and helping you improve the accuracy of your label.

This study was approved by the Institutional Review Board (IRB) at CMU. We will not identify you, your app, or your organization in any publications that come out of this research without your written permission.

Is that all OK? If yes, please sign the consent form.

Is it OK if I record the interview? [Start recording after receiving their positive answer]

\subsection{Background Questions}

Now I’ll introduce some background about the Google data safety section, which is the main topic we’re discussing today. I prepared some slides and I’m gonna share my screen.

(The slides contain screenshots of Google Play safety labels. After showing the slides, ask the following questions)

\begin{itemize}
    \item Have you heard about them before?
    \item Have you created any of these labels before?
    \item Have you heard about the iOS privacy label before? If so, have you created it for any iOS apps?
\end{itemize}

Before we get started, I’d love to learn a little bit more about your app. Can you briefly tell me:

\begin{itemize}
    \item What’s the app designed for?
    \item What was your role in the development?
    \item Is it still under active development?
\end{itemize}

\subsection{Task 1: Use Google Play Developer Console to Create the Label}

Verbal instruction: Now I’ll introduce today’s first task. I’d like you to log into the google play developer console using a test account provided by us, and create a label of the selected app. The label should accurately represent the data practices of the selected app.

Please handle this task as you normally would and take as long as you need. You are welcome to look at any documentation you would normally consult, except for the app's privacy label if it’s available. In order for us to see any resources you use, please either share your full screen or open any additional resources in the same window where you’re completing the task. 

If you need a resource that is not currently available or would ordinarily ask somebody for help, please say aloud what resources you would use and who you would usually contact.

Please try to keep thinking aloud during this process. Basically that means tell me whatever comes to your mind when working on this task, such as say your thought process aloud or voice any questions or comments you have. When you think you’re done, just let me know.

\subsection{Task 2: Use Matcha to Create the Label}

Verbal instruction: Our lab developed an Android Studio plugin to help you create the safety label in a semi-automated way. In the second task, you will create the label again using this tool. Now let me help you install the plugin and set up the environment.

[After installing the plugin] I have prepared a short video introducing how to use this plugin. I’ll send the link to you. Could you play it from your end? This video contains some sound. Please let me know if the sound doesn’t work correctly on your end.

[After the tutorial] Do you have any questions?

[After answering their questions] Now let’s go back to the IDE. Similar to the first task, please try to keep thinking aloud during this process. Basically that means tell me whatever comes to your mind when working on this task, such as say your thought process aloud or voice any questions or comments you have. 

\subsection{Post-Study Interview}

Now I would like to compare the two safety labels that you just created. For the first label you just created on the developer console, please open the developer console to show the preview. For the second label created with Matcha, please switch to the “label preview” tab in the IDE plugin. Put the two windows side by side so we can compare the results. We’re anticipating there may be some discrepancies.

Before we compare the results, I want to reassure you that the goal of this study is not to measure your ability, and discussing these discrepancies will help us understand the effectiveness of our tool and identify challenges developers may encounter when handling this task, so please don’t be shy in noting any inaccuracies in either label. Your perspective is really helpful, and no identifying information will be shared about you, your app, or your company, in our report.

When I go through each discrepancy instance between the two versions, could you tell me;
\begin{itemize}
    \item Which version do you think is more accurate?
    \item What do you think could possibly cause the difference between the two privacy labels?
\end{itemize}

Next, I'd like to ask a few questions about your experience using the two tools.

How is the experience of using the Google Play developer console and Matcha? Which one do you prefer? Why?

What do you think about using Matcha to generate privacy labels in general?

[Showing the key features of Matcha using a few slides] Could you tell me which are the three features of Matcha that you felt the most useful and why?

We hope to deploy this tool in the future and would like feedback to help us improve the design and implementation. Is there anything that we can improve in this tool that can make you more likely to install and use it?
Any other thoughts to share?
We’ll continue working on Matcha, if you have any friends that might be interested in it, let us know!

\section{Participant Overview}
\label{sec:participant-overview}

\autoref{tab:particpant-overview} provides a detailed overview of the background of each participant and their app selected for the study.

\begin{table*}[h]
    \centering
    \caption{Participant Overview. Our sample features a good sample of developers and apps across several dimensions, including participant's geographic location (\textit{Location}), app development purpose (\textit{Purpose}), app development team size (\textit{Team Size}), app downloads (\textit{Downloads}), the current data safety label on Google Play (Current label), and participant's role(s) in the development team (\textit{Participant's Role(s) In Team}). Nine out of the 12 participants had prior experience in publishing apps on the Google Play store (Play). The app development purposes involve four options, covering situations when the participant developed the app as part of their job (\textit{Job}), as part of their hobby (\textit{Hobby}), for a course project (\textit{Course}), and for a research project (\textit{Research}).}
    \begin{tabular}{p{0.02\linewidth} p{0.04\linewidth} p{0.1\linewidth} p{0.07\linewidth} p{0.04\linewidth} p{0.09\linewidth}  p{0.25\linewidth} p{0.25\linewidth}}
    \toprule
    ID & Play & Location &  Purpose  & Team Size & Downloads &  Current label & Participant's Role(s) in Team \\
    \midrule
    \uten{} & yes & Pakistan & Job & 2-5  & 1M+ & No data shared, 4 data types in 3 categories collected (App activity, App info and performance, and Device or other IDs) & Mobile App Developer,
Designer\\
    \efour{} & no & U.S.  & Course & 2-5 & Not Play & N/A & Mobile App \& Backend Developer, Designer\\
    \esix{} & no & Nigeria & Hobby & 1 & Not Play & N/A & Mobile App \& Backend Developer, Designer, Project Manager \\
    \fone{} & yes & Ukraine & Hobby & 1 & Not Play & N/A & Mobile App Developer\\
    \ftwo{} & yes & Georgia & Job & 2-5 & 50K+ & No data shared, 6 data types in 4 categories collected (Personal info, Photos and videos, App activity, and Device or other IDs) & Mobile App Developer\\
    \eeight{} & no & U.S. & Course & 2-5 & Not Play & N/A & Mobile App \& Backend Developer \\
    \ffive{} & yes & Pakistan & Job & 1 & Not Play & N/A & Mobile App Developer\\
    \fsix{} & yes & Pakistan & Job & 1 & 100+ & No data shared, no data collected & Mobile App Developer\\
    \fthree{} & yes & Bangladesh & Hobby & 1 & 500+ & Not provided & Mobile App Developer \\
    \fseven{} & yes & Egypt & Course & 1 & Not Play & N/A & Mobile App Developer\\
    \fnine{} & yes & Pakistan & Job & 1 & 100K+ & Not provided & Mobile App Developer\\
    \feleven{} & yes & India & Job & 1 & 100K & Not provided & Mobile App Developer \\
    \bottomrule
    \end{tabular}
    \label{tab:particpant-overview}
\end{table*}

\section{Qualitative Analysis Code book}
\label{sec:codebook}

We present the final code book of our qualitative analysis in \autoref{tab:codebook}.

\begin{table*}[]
    \centering
    \caption{The complete codebook of our qualitative analysis of the interview recordings}
    \begin{tabular}{p{0.08\linewidth} p{0.25\linewidth} p{0.3\linewidth} p{0.3\linewidth}}
    \toprule
    Theme & Code & Memo & Example \\
    \midrule
    Cause of error & Forgetfulness & The developer mentioned they forgot something about their apps, such as libraries integrated in the app or a feature implemented in the app. & \blockquote[\ffive{}]{Sorry, I forgot about this section. It was for users to add text to their photos} \\
    & Library & The developer mentioned misunderstanding related to third-party libraries. & \blockquote[\efour{}]{This one is more precise, because I did not know that the library was was doing it on its own behind the screen. So that's why I did not put this information.} \\
    & Misunderstanding about the task & The developer mentioned they did not understand something related to the data safety label creation task. & \blockquote[\efour{}]{I did not know that I need to report out other user generated content.} \\
    & lack technical knowledge & The developer mentioned something that demonstrated their misunderstanding about technical concepts. & \blockquote[\efour{}]{I'm not sure if that is if I'm actually using the precise location are like an approximate location. That's where I'm confused.} \\
    \midrule
    Comment on Matcha & prefer Matcha - informative & The developer mentioned Matcha helped them learn useful information. & \blockquote[\ftwo{}]{I never care about data collection and I don't even look at what we do. So I think I learned a lot from this} \\
    & prefer Matcha - better flexibility & The developer mentioned Matcha gave them better flexibility in the label creation process. & \blockquote[\esix{}]{I want it because it gives me the flexibility and it give me the feeling of a developer's mindset.} \\
    & prefer Matcha - better engagement & The developer mentioned Matcha better involved them in the task. & \blockquote[\efour{}]{It's a lot more involved process when you're using Matcha than Google Play console.} \\
    & prefer Matcha - accuracy & The developer mentioned Matcha improved the label accuracy. & \blockquote[\fone{}]{I prefer the plugin because it can search the privacy leak for developers.} \\
    & prefer Matcha - easy to use & The developer mentioned Matcha was easy to learn and use. & \blockquote[\eeight{}]{I think the quickfix for the annotation is pretty convenient} \\
    & issue - redundancy & The developer complained that several tasks felt repetitive and unnecessary to them. & \blockquote[\eeight{}]{Like for some strange reason I think, it looks for the word search, but isn't search really common?} \\
    \bottomrule
    \end{tabular}
    \label{tab:codebook}
\end{table*}

\end{document}